\documentclass{article}

\usepackage{geometry}

\usepackage{tikz}
\usetikzlibrary{positioning}

\usepackage{amsmath}

\usepackage[utf8]{inputenc} 
\usepackage[T1]{fontenc}    
\usepackage{hyperref}       
\usepackage{url}            
\usepackage{booktabs}       
\usepackage{amsfonts}       
\usepackage{nicefrac}       
\usepackage{microtype}      
\usepackage{lipsum}
\usepackage{graphicx}
\graphicspath{ {./figs/} }

\usepackage{amsthm}
\usepackage{hyperref}

\newcommand{\dom}{\mathbf{dom}}
\newcommand{\OD}{\mathrm{OD}}

\newcommand{\V}{\mathcal{V}}
\newcommand{\Vin}{\mathcal{V}_{\rm in}}
\newcommand{\Vout}{\mathcal{V}_{\rm out}}
\newcommand{\ru}{\mathbf{r}}
\newcommand{\Prog}{\mathbf{P}}
\newcommand{\SP}{\mathit{sub}}
\newcommand{\paths}{\mathit{paths}}
\newcommand{\CC}[2]{\mathit{concat}{(#1,#2)}}
\newcommand{\F}{\mathcal{F}}
\newcommand{\PF}{\mathcal{PF}}

\newcommand{\new}[1]{#1}

\theoremstyle{remark}
\newtheorem{remark}{Remark}[section]
\newtheorem{example}[remark]{Example}
\theoremstyle{plain}
\newtheorem{proposition}[remark]{Proposition}
\newtheorem{theorem}[remark]{Theorem}
\newtheorem{lemma}[remark]{Lemma}

\title{J-Logic: \new{a Logic for Querying JSON}}

\author{
 Jan Hidders \\
  University of London\\
   \and
 Jan Paredaens \\
  Universiteit Antwerpen \\
  \and
 Jan Van den Bussche \\
  Universiteit Hasselt
}

\date{}

\begin{document}

\maketitle

\begin{abstract}
We propose a logical framework, based on Datalog, to study the
foundations of querying JSON data.  The main feature of our
approach, which we call J-Logic, is the emphasis on paths.  Paths
are sequences of keys and are used to access the tree structure
of nested JSON objects.  J-Logic also features ``packing'' as a
means to generate a new key from a path or subpath.  J-Logic with
recursion is computationally complete, but many queries can be
expressed without recursion, such as deep equality.
We give a necessary condition for
queries to be expressible without recursion.  Most of our results
focus on the deterministic nature of JSON objects as partial
functions from keys to values.  Predicates defined by J-Logic
programs may not properly describe objects, however.
Nevertheless we show that every object-to-object transformation
in J-Logic can be defined using only objects
in intermediate results.  Moreover we show that it is decidable
whether a positive, nonrecursive J-Logic program always returns
an object when given objects
as inputs.  Regarding packing, we show that packing is
unnecessary if the output does not require new keys.  Finally, we
show the decidability of query containment for positive,
nonrecursive J-Logic programs.

This papter is the extended version of an earlier version
published in the proceedings of SIGMOD/PODS 2017  \cite{Hidders:2017}.
\end{abstract}

\section{Introduction}

JSON is a popular semistructured data model used in NoSQL systems
and also integrated in relational systems.  Proposals for
expressive query languages for JSON include JSONiq
\cite{florescu_jsoniq,jsoniq_book},
which is based on XQuery, and SQL++ \cite{sql++}, which is based
on SQL\@.  Schema formalisms for JSON are also being investigated
\cite{chili_jsonschema}.  Hence the time is ripe to investigate the
logical foundations of JSON querying, which is the goal of the
present paper.  

A JSON object is a partial function, mapping keys to
values.  Here, a value is either an atomic value or an
object in turn.  Hence, objects can be nested, and thus can
be viewed as trees, similarly to XML documents.  JSON trees have
some special characteristics, however, which form the starting
point of our work.  A first
difference with XML trees is that JSON trees are edge-labeled rather than
node-labeled; the keys are the edge labels. More importantly,
JSON trees are \emph{deterministic} in the sense of Buneman,
Deutsch and Tan
\cite{buneman_deterministic,tajima_bunemandeterministic}.
Specifically, since objects are functions, different edges
from a common parent must have different
labels.\footnote{Buneman, Deutsch and Tan actually considered an
extension of JSON where keys need not be atomic, but can be 
objects in turn.}

Determinism is convenient because paths starting in the root of a
given tree can be identified with key sequences.\footnote{In JSON
Schema \cite{chili_jsonschema}, key sequences are called ``JSON
pointers''.} This suggests an alternative view of objects as sets
of path--value pairs, where each path is a path from the root to
a leaf, and the corresponding value is the atomic value of that
leaf.  We call such a set of path--value pairs an \emph{object
description}.  Since paths are sequences of keys, we are led to
the conclusion that to query JSON objects, we need a query
language that can work with sets of sequences.

At the same time, the theory of query languages is solidly
grounded in logic \cite{ahv_book}. Datalog in particular is a
convenient logic-based language with a long tradition in data
management research and a wide variety of current applications
\cite{datalog_sigmod2011,datalogreloaded,datalog2.0,vadalog_linguafranca}.

We are thus motivated to investigate the logical foundations for
JSON querying within a Datalog language for sets of sequences.
Such a language, called sequence Datalog, has already been introduced by
Bonner and Mecca
\cite{bonnermecca_sequences,meccabonner_termination,bonnermecca_transducers}.
Bonner and Mecca were primarily interested in expressive sequence
manipulation, of the kind needed in bioinformatics applications.
They reported results on expressiveness, complexity of
computations, and on ways to combine recursion with sequence
concatenation while still guaranteeing termination or
tractability.

In this paper, we focus more on questions motivated by
JSON querying and deterministic semistructured data.  Thereto, we
propose a new approach \new{based on} sequence Datalog, called J-Logic.
Moreover, J-Logic adds a feature for \new{constructing} new keys,
called \emph{packing}.  Key generation is necessary if we want
the result of a query over objects to be again an object.  Consider,
for example, the Cartesian product of two objects that have
$N$ keys each.  The result needs to be a object with $N^2$
keys. So, we cannot manage by just reusing the keys from the
input; new keys must be generated.

\begin{figure*}
\begin{center}
\begin{minipage}{0pt}
\begin{tabbing}
$T(\langle @x.@y \rangle . r . @x . \$x' : @u) \gets
R(@x.\$x':@u), S(@y.\$y':@v)$
\\
$T(\langle @x.@y \rangle . s . @y . \$y' : @v) \gets
R(@x.\$x':@u), S(@y.\$y':@v)$ 
\end{tabbing}
\end{minipage}
\end{center}
\caption{\mdseries J-Logic program defining $T$ as the
Cartesian product of $R$ and $S$.  Here, $@x$ and $@y$ are atomic
variables, binding to the top-level atomic keys of $R$ and $S$
respectively; $\$x'$ and $\$y'$ are path variables, binding to
the paths in the subobjects below $@x$ and $@y$ in $R$ and $S$
respectively. The variables $@u$ and $@v$ bind to atomic values
stored in the leaves.  The dot indicates concatenation.  We also
use constant keys $r$ and $s$ to indicate the $R$- and $S$-parts
of each pair of the Cartesian product.}
\label{figcart}
\end{figure*}

The creation of new data elements (keys, identifiers, nodes, and
so on) in the result of a query has already been considered in
many contexts, such as highly expressive languages
\cite{av_proc,av_datalog}, object databases
\cite{ak_iql,good_ieee,kv_ldm}, information integration
\cite{tsimmis}, data exchange \cite{miller_invention}, and
ontology based data access \cite{obda}.  The popular languages
XQuery and SPARQL both have node creation.  In logic based
approaches, element creation is typically achieved through the
use of Skolem functions \cite{hy_ilog,kw_pods89}.

In J-Logic, however, we can take advantage of having
sequences in the language.  We can
generate new keys simply by \emph{packing} a key
sequence $s$ into a new key $\langle s \rangle$.  For example,
consider two objects
\begin{quote}
$R = \{ a : o_1, b : o_2 \}$ and
$S = \{ c : o_3, d : o_4 \}$,
\end{quote}
where $o_1$, $o_2$, $o_3$, and $o_4$ are subobjects.
We can represent the Cartesian product of $R$ and $S$ by the object
$$
\begin{aligned}
T = \{&\langle a.c \rangle : \{r:\{a:o_1\},s:\{c:o_3\}\},\\
&\langle a.d \rangle : \{r:\{a:o_1\},s:\{d:o_4\}\},\\
&\langle b.c \rangle : \{r:\{b:o_2\},s:\{c:o_3\}\},\\
&\langle b.d \rangle : \{r:\{b:o_2\},s:\{d:o_4\}\}\}.
\end{aligned}
$$
The two J-Logic rules in Figure~\ref{figcart} accomplish this.

Packed keys should be seen as an intermediate construct.
We envisage that any packed keys
present in the final result of a query will be replaced by fresh
identifiers, as in the ILOG approach \cite{hy_ilog}.
For example, $T$ above could be returned in the
following form:
$$
\begin{aligned}
T = \{&t_1 : \{r:\{a:o_1\},s:\{c:o_3\}\},\\
&t_2 : \{r:\{a:o_1\},s:\{d:o_4\}\},\\
&t_3 : \{r:\{b:o_2\},s:\{c:o_3\}\},\\
&t_4 : \{r:\{b:o_2\},s:\{d:o_4\}\}\}.
\end{aligned}
$$

The aforementioned languages SQL++ and JSONiq do not have key
generation: there, the Cartesian product can be computed as a bag
(or sequence) of objects, but not as one object itself.  Key
generation can thus be seen as an alternative to adding an extra
collection feature (like bags, or sequences if we agree on some
way to order objects) to the query language.  We admit that a bag
of objects could be easily transformed into one object by
generating fresh keys.  Thus the two approaches (key generation, or
bags that are eventually transformed into objects) are largely
equivalent.  In J-Logic we have chosen for key generation through
packing, because it is a lightweight addition to sequence
Datalog. Moreover, it allows us to work with just a single kind
of collections, namely, objects (more precisely, object
descriptions).

In this paper we will show the following results.
\begin{enumerate}
\item
J-Logic programs may be recursive, but we are mostly interested
in the nonrecursive case.  Nonrecursive programs have
polynomial-time data complexity, and due to the use of sequence
variables, nonrecursive programs are already quite powerful.
We give a necessary condition on queries computable by
nonrecursive programs, which can serve as a tool to show that
certain queries involving objects of unbounded depth require recursion.
(Nonrecursive J-Logic over objects of bounded depth is
essentially equivalent to relational algebra.)
\item
We show the technical result that packing,
while convenient and necessary in general,
is not needed to compute queries from flat inputs to flat
outputs.  Here, \emph{flat} means that no packed keys occur in the data.
An open question is whether this can be done without
recursion (our simulation of packing needs recursion).  An affirmative answer
would yield a result analogous to
the ``flat--flat theorem'' for the
nested relational algebra \cite{pvg_flatflat} or calculus
\cite{bntw}.
\item
In J-Logic, a JSON object is described as a mapping from
root-to-leaf paths to atomic values.  Accordingly, predicates defined by
J-Logic rules are relations between paths and atomic values.  Not
every such relation properly describes a JSON object, however.
Nevertheless, we show the ``object--object theorem'': every
query from objects to objects, computable by a J-Logic
program, is computable by a J-Logic program so that every
intermediate relation is a proper object description.
\item \label{result4}
The object--object theorem assumes a J-Logic program that maps
objects to objects.  But can we check this?
We show that the object--object property is decidable
for positive, nonrecursive programs.  We do this by adapting the
chase procedure for equality-generating dependencies,
well known from relational databases
\cite{ahv_book}.  In our model, however, the chase is not complete in general.
We nevertheless can use it resolve our problem.
\item \label{result5}
Finally, we show that the containment problem for
positive, nonrecursive programs, over flat instances, is decidable.
To the best of our knowledge, the
containment problem was not yet addressed in the setting of
sequence Datalog.  We solve the
problem in our setting by extending the known inclusion test for pattern
languages over an infinite alphabet \cite{fr_patterninclusion}.
\end{enumerate}

This paper is further organized as follows.  In
Section~\ref{secmodel} we introduce our formalization of the JSON
data model.  In Section~\ref{seclanguage} we define J-Logic.
In Section~\ref{secexpr} we discuss the expressive power of
nonrecursive J-Logic and state the flat--flat theorem.
In Section~\ref{secoo} we discuss the problem of proper object
descriptions, state the object--object theorem, and study the
object--object decision problem.
Section~\ref{secontain} is devoted to the containment
problem.  We conclude in Section~\ref{seconc}.

\section{A formal data model based on JSON} \label{secmodel}

We begin by defining our formalization of the JSON data model.
From the outset we assume an infinite domain $\dom$ of
atomic data elements, which we call \emph{atomic keys}.  In practice,
these would be strings, numbers, or any other type of data that the
database system treats as atomic.  Now the sets of \emph{values} and
\emph{objects} are defined as the smallest sets satisfying the
following:
\begin{itemize}
\item
Every atomic key is a value;
\item
Every object is a value;
\item
Every mapping from a finite set of atomic keys to values is an
object.
\end{itemize}
Recall that a mapping is a set of pairs where no two pairs have the
same first component.  Thus, an object is a set of key--value pairs.
It is customary to write a key--value pair
$(k,v)$ in the form $k:v$.  For an object $o$ and a key $a$, we
sometimes use the notation $o.a$ for the $a$-value of $o$, i.e.,
for $o(a)$.

\begin{example}
\label{exobject}
Using strings such as `name', `age', `anne', `bob' and `chris',
and numbers such as $12$, $18$ and $24$, as atomic keys, the following are
three examples of objects:
$$ \begin{aligned}
o_1 &= \{\mathrm{name}:\mathrm{anne},\mathrm{age}:12\} \\
o_2 &= \{\mathrm{name}:\mathrm{bob},\mathrm{age}:18\} \\
o_3 &= \{\mathrm{name}:\mathrm{chris},\mathrm{age}:24\}
\end{aligned} $$
Since objects can be nested, the following is also an object:
$$ o=\{\mathrm{name}:\mathrm{john},
\mathrm{children}:\{1:o_1,2:o_2,3:o_3\}\} $$
We have $o.\mathrm{children}.2=o_2$.
Finally, note that the set
$\{\mathrm{name}:\mathrm{anne},\mathrm{name}:\mathrm{bob}\}$ is
not an object since it is not a well-defined mapping.  The set
$\{\mathrm{anne}:\mathrm{name},\mathrm{bob}:\mathrm{name}\}$,
however, is perfectly allowed as an object.\qed
\end{example}

\begin{remark}
Some remarks are in order.
\begin{enumerate}
\item
In the JSON standard \cite{jsonstandard}, the keys in an object
can only be strings, but values can be numbers.
In our formalization we make no distinction between different
types of atomic data, which explains the example above where we
used the numbers $1$, $2$ and $3$ as keys.
In the language JavaScript, an array may be viewed
as an object with numbers as keys.  So, our approach is not
too much at odds with reality.
\item
Indeed, the JSON standard also has arrays besides objects.  In this paper
we focus on unordered objects.  An extension of our approach, where a total
order is assumed on atomic keys (and extended to packed keys, see
later) seems feasible and would be able to model arrays.
\item
The term ``atomic \emph{key}'' is a
bit misleading, as these elements may not only be used as keys,
but also as values.  Indeed, that keys can occur as data values, and vice
versa, is a characteristic feature of JSON\@.
\end{enumerate}
\end{remark}

\paragraph{Packed keys}
Until now
we have defined an object as a mapping from atomic keys to values.  Since these
values can be objects in turn, we can use sequences of atomic keys to
navigate deeper inside an object.
Sequences of keys will be called paths.
Moreover, we also introduce
packed keys, as they can be created by J-Logic rules.
Formally, the sets of \emph{keys} and
\emph{paths} are defined as the smallest sets such that
\begin{itemize}
\item
every atomic key is a key;
\item
if $p$ is a path then $\langle p \rangle$ is a key, called a
\emph{packed key};
\item
every nonempty finite sequence of keys is a path.
\end{itemize}
In our notation, we use dots to separate the elements of a
sequence.  At the same time, the dot will be used to denote
concatenation of paths.

\begin{example}
Let $a$ and $b$ be atomic keys.  Then $a.b$ is a path; $k=\langle
a.b \rangle$ is a packed key; $p= b.b.k.a$ is again a path; and
$\langle p \rangle$ is again a packed key.\qed
\end{example}

From now on we allow packed keys in objects.  Thereto we
generalize the notion of object by defining an object to be
a mapping from a finite set of keys to values.  Thus,
keys need not be atomic but can also be packed.  We
already saw an example of a object with packed keys, $T$ in the
Introduction.

\paragraph{Object descriptions}  An object can be visualized as a
tree, where edges are labeled with keys and leaves are labeled
with \emph{atomic values}: atomic keys or $\emptyset$ (the empty
object).  Thus, we can completely describe an object by listing all paths
from the root to the leaves, and, for each such path, giving the
label of the corresponding leaf.

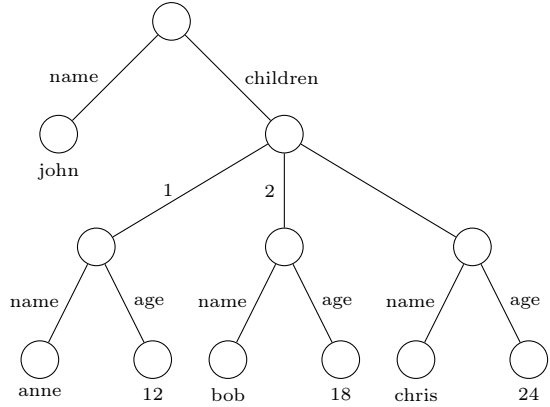
\begin{figure}
\centering
\begin{tikzpicture}[
    level 1/.style={sibling distance=3cm},
    level 2/.style={sibling distance=2.5cm},
    level 3/.style={sibling distance=1.5cm},
    level distance=1.5cm,
    every node/.style = {font=\scriptsize},
    treenode/.style = {shape=circle, minimum width=0.5cm, minimum height=0.5cm,
    draw, align=center}]
  \node [treenode] {}
    child { node  [treenode, label={below:{john}}] {} 
      edge from parent node[left=0.1cm, draw=none] {name} }
    child { node [treenode] {} 
      child { node [treenode] {}
        child { node [treenode, label={below:{anne}}] {} 
          edge from parent node[left=0cm, draw=none] {name} }
        child { node [treenode, label={below:{12}}] {} 
          edge from parent node[right=0cm, draw=none] {age} }
        edge from parent node[left=0.1cm, draw=none] {1} }
      child { node [treenode] {}
        child { node [treenode, label={below:{bob}}] {} 
          edge from parent node[left=0cm, draw=none] {name} }
        child { node [treenode, label={below:{18}}] {} 
          edge from parent node[right=0cm, draw=none] {age} }
        edge from parent node[left, draw=none] {2} }
      child { node [treenode] {}
        child { node [treenode, label={below:{chris}}] {} 
          edge from parent node[left=0cm, draw=none] {name} }
        child { node [treenode, label={below:{24}}] {} 
          edge from parent node[right=0cm, draw=none] {age} }
      }
      edge from parent node[right=0.1cm, draw=none] {children}
    };
\end{tikzpicture}
\caption{\mdseries Object $o$ from Example~\ref{exobject} as a tree.}
\label{figtree}
\end{figure}

\begin{example}
Recall the object $o$ from
Example~\ref{exobject}.  Figure~\ref{figtree} shows $o$
as a tree.  Its description as a set of
path--value pairs is as follows:
$$ \begin{aligned}
& \rm name:john \\
& \rm children.1.name:anne \\
& \rm children.1.age:12 \\
& \rm children.2.name:bob \\
& \rm children.2.age:18 \\
& \rm children.3.name:chris \\
& \rm children.3.age:24
\end{aligned} $$
\nopagebreak
\qed
\end{example}

Formally, we define an \emph{object description} to be any set of
pairs of the form $p:v$, where $p$ is a path and $v$ is an atomic
value.  If $o$ is an object, the object description of $o$,
denoted by $\OD(o)$, is defined inductively as follows:
\begin{itemize}
\item
If $o$ is $\emptyset$, or a singleton object of the form
$\{k:b\}$ with $b$ an atomic value, then $\OD(o) = o$.
\item
If $o$ is a singleton object of the form $\{k:o'\}$, with $o'$
an object, then $$ \OD(o) = \{k.p:b \mid (p:b) \in \OD(o')\}. $$
\item
If $o$ is a non-singleton object, then $$ \OD(o) = \bigcup \{
\OD(\{k:v\}) \mid (k:v) \in o\}. $$
\end{itemize}

\begin{remark} \label{remproper}
Not every finite object description is the object description of some
object; those that are, are called \emph{proper}.  Simple examples
of improper object descriptions are $\{a:1,a:2\}$ and $\{a:1,a.a:1\}$.
We will focus on proper object descriptions in Section~\ref{secoo}.
For now, we allow arbitrary object descriptions.
\end{remark}

\paragraph{Vocabularies, instances, and queries}

We can finally define the fundamental notions of data\-base
instance and query in our data model.  Just like a relational
database instance is a finite collection of named relations, here
we will define an instance as a finite collection of named object
descriptions.  Since object descriptions are binary relations
(sets of pairs), we refer to their names as ``relation names''.

Formally, a \emph{vocabulary} $\V$ is a finite set of relation
names.  An \emph{instance} $I$ over $\V$ assigns to each name $R
\in \V$ an object description $I(R)$.  Given two disjoint vocabularies
$\Vin$ and $\Vout$, a \emph{query from $\Vin$ to $\Vout$} is a
partial function from instances over $\Vin$ to instances over
$\Vout$.

In database theory one often focuses on \emph{generic} queries
\cite{ahv_book}.  We can define a similar notion of genericity
here.  Let $f$ be a permutation of $\dom$.  Then $f$ can be
extended to paths, packed keys, object descriptions, and
instances, simply by applying $f$ to every occurrence of an
atomic key.  Let $C$ be a finite subset of $\dom$ (these are the
atomic keys that would be explicitly mentioned in a program
for the query).  Then a query $Q$ is called \emph{$C$-generic} if
for every permutation $f$ of $\dom$ that is the identity on $C$,
and for every instance $I$, we have $Q(f(I)) = f(Q(I))$.  In
particular, if $Q(I)$ is undefined, then $Q(f(I))$ must also be
undefined.

\section{J-Logic} \label{seclanguage}

In the syntax of J-Logic, we assume disjoint supplies of
\emph{atomic variables} (ranging over atomic keys) and \emph{path
variables} (ranging over paths).  The set of all variables is
also disjoint from $\dom$.  We indicate atomic variables as $@x$
and path variables as $\$x$.

\emph{Key expressions} and \emph{path expressions} are defined
just like keys and paths, but with variables added in.
Formally, we define
the sets of key expressions and path expressions
to be the smallest sets such that
\begin{itemize}
\item
every atomic key is a key expression, called a \emph{constant};
\item
every atomic variable is a key expression; constants and atomic
variables are also called \emph{atomic key expressions};
\item
if $e$ is a path expression then $\langle e \rangle$ is a key
expression, called a \emph{packed key expression};
\item
every nonempty finite sequence of key expressions and path
variables is a path expression.
\end{itemize}

Recall that an atomic value is an atomic key or $\emptyset$.  Now
an \emph{atomic term} is an atomic value or an atomic variable.

A \emph{predicate} is an expression of the form $P(e:t)$, with
$P$ a relation name, $e$ a path expression, and $t$ an atomic term.

An \emph{equality} is an expression of the form $e_1 = e_2$, with
$e_1$ and $e_2$ path expressions.

Many of the following definitions adapt the standard definition
of Datalog \cite{ahv_book} to our data model.

An \emph{atom} is a predicate or an equality.  A \emph{negated
atom} is an expression of the form $\neg A$ with $A$ an
atom.  A \emph{literal} is an atom (also called a positive
literal) or a negated atom (a negative literal).

A \emph{body}
is a finite set of literals.

A \emph{rule} is an expression of the form $H \gets B$, where $H$
is a predicate, called the \emph{head} of the rule, and $B$ is a body.
We define the \emph{limited variables} of the rule as the
smallest set such that
\begin{itemize}
\item
every variable occurring in a positive predicate in $B$ is limited;
and
\item
if all variables occurring in one of the sides of a positive
equality in $B$ are limited, then all variables occurring in
the other side are also limited.
\end{itemize}
A rule is called \emph{safe} if all variables occurring in the
rule are limited.

Finally, a \emph{program} is a finite set of safe rules with
stratified negation.  We omit the definition of stratified
negation, which is well known \cite{ahv_book}.  For our purposes
in this paper, stratified negation suffices.  A program is called
\emph{positive} if it does not use negation.  We also assume
familarity with the distinction between recursive and
nonrecursive programs.

\paragraph{Semantics}

We have defined the notion of instance as an assignment of object
descriptions to relation names.  A convenient equivalent view of
instances is as sets of \emph{facts}.  A fact is an
expression of the form $R(p:v)$ with $R$ a relation name, $p$ a
path, and $v$ an atomic value.  An instance $I$ over
vocabulary $\V$ is viewed as the set of facts $$ I = \{R(p:v)
\mid R \in \V \text{ and } (p:v) \in I(R)\}. $$

A \emph{valuation} is a function $\nu$ defined on a finite set of
variables, that maps atomic variables to atomic keys and path
variables to paths.  We say that $\nu$ is \emph{appropriate} for
a syntactical construct (such as a path expression, a
literal, or a rule) if $\nu$ is defined on all variables
occurring in the construct.  We can apply an appropriate valuation $\nu$ to
a key or path expression $e$ in the obvious manner: we substitute
each variable by its image under $\nu$ and obtain a key or a path
$\nu(e)$.  Likewise, we can apply an appropriate valuation to a
predicate and obtain a fact.

Let $L$ be a literal, $\nu$ be a valuation appropriate for $L$,
and $I$ be an instance.  The definition of when $I,\nu$
\emph{satisfies} $L$ is as expected: if $L$ is a predicate, then
the fact $\nu(L)$ must be in $I$; if $L$ is an equality
$e_1=e_2$, then $\nu(e_1)$ and $\nu(e_2)$ must be the same path.
If $L$ is a negated atom $\neg A$, then $I,\nu$ must not
satisfy $A$.

A body $B$ is satisfied by $I,\nu$ if all its literals are.  Now
a rule $\ru = H \gets B$ is satisfied in $I$ if for every
valuation $\nu$ appropriate for $\ru$ such that $I,\nu$ satisfies
$B$, also $I,\nu$ satisfies $H$.

The notions of EDB and IDB relation names of a program are well
known: the IDB relation names are the relation names used in the
head of some rules; the other relation names are the EDB relation
names.  Given a vocabulary $\Vin$, a program is said to be
\emph{over $\Vin$} if all its EDB relation names belong to
$\Vin$, and its IDB relation names do not.

Now the semantics of programs with stratified negation is defined
as usual \cite{ahv_book}.  Recall that a program is called
semipositive if negative predicates only use EDB relation names.
We first apply the first stratum, which is semipositive, and then
apply each subsequent stratum as a semipositive program to the
result of the previous stratum.  So we only need to give
semantics for semipositive programs.

Let $\Prog$ be a semipositive program
over $\Vin$, and let $I$ be an instance over $\Vin$.  Let $\V$ be
the set of IDB relation names of $\Prog$.  Then $\Prog(I)$ is the
smallest instance over $\Vin \cup \V$ that satisfies all the
rules of $\Prog$, and that agrees with $I$ on $\Vin$.

In the end, a program $\Prog$ over $\Vin$ can be used to compute a query
$Q$ from $\Vin$ to $\Vout$, for any designated subset $\Vout$ of the
IDB relation names of $\Prog$.  Here, $Q(I)$ simply equals
the restriction of $\Prog(I)$ to $\Vout$.

\paragraph{Syntactic sugar}

We have kept the syntax of J-Logic minimal so as to keep the
formal definitions as simple as possible.  For writing practical
programs, however, it is convenient to introduce some syntactic sugar:
\begin{itemize}
\item
Variables of the form $\%u$ range over atomic values, i.e.,
atomic keys or $\emptyset$.  We could always eliminate such a variable in
a rule by splitting the rule in two: one in which we replace
$\%u$ by a normal atomic variable $@u$, and one in which we
replace $\%u$ by $\emptyset$ (and resolve equalities accordingly).
\item
Variables of the
form $?z$ range over paths or the empty sequence (recall that
paths are nonempty).  As long as such a variable is only used
concatenated with other path expressions, we could always
eliminate it from a rule by splitting the rule in two: one in
which we replace $?z$ by a normal path variable $\$z$, and one in
which we simply delete all occurrences of $?z$ (resolving
equalities accordingly).
\item
Variables of the form $\#z$ range over keys, atomic as well as
packed.  We could always eliminate such a variable in a rule by
splitting the rule in one where we replace $\#z$ by
an atomic variable $@z$, and one where we replace $\#z$ by a
packed key expression $\langle \$z \rangle$.
\end{itemize}

\paragraph{Examples}

We aim to illustrate that J-Logic does not need recursion to
express many useful queries involving deeply nested data.  We
begin, however, by illustrating why nonrecursive programs are
desirable.

\begin{example}[Nontermination]
Due to the use of concatenation in heads of rules,
the result of a recursive program applied to a finite instance
may be infinite.  A simple example is the following: (this
program has no EDB relation names; the body of the first rule is
empty)
\begin{tabbing}
$S(a:\emptyset) \gets {}$ \\
$S(a.\$x:\emptyset) \gets S(\$x:\emptyset)$
\end{tabbing}
We consider such programs to be nonterminating.  For limited
forms of recursion that guarantee termination or even
tractability, we refer to the work of Bonner and Mecca
\cite{bonnermecca_sequences,meccabonner_termination}.
Nonrecursive programs clearly always terminate.\qed
\end{example}

\begin{example}[Deep equality]
The following nonrecursive program is applied to the object
description $R$ of an object $o$, assumed to have values $o.a$
and $o.b$.  The program tests equality of $o.a$ and
$o.b$; if so, it outputs the fact $Q(\mathrm{yes}:\emptyset)$
\new{and if not, it outputs no facts}.
Note that $o.a$ and $o.b$ may be atomic keys
(case handled by the first and last rule), or may be
objects themselves.  Thus, the other rules of the program
test set equality of the object descriptions of $o.a$
and $o.b$.
\begin{tabbing}
$T(\mathrm{atomic}:\emptyset) \gets R(a:\%u), R(b:\%v)$ \\
$Q'(\mathrm{no}:\emptyset) \gets R(a.\$x:\%u), \neg R(b.\$x:\%u)$ \\
$Q'(\mathrm{no}:\emptyset) \gets R(b.\$x:\%u), \neg R(a.\$x:\%u)$ \\
$Q(\mathrm{yes}:\emptyset) \gets \neg
T(\mathrm{atomic}:\emptyset), \neg Q'(\mathrm{no}:\emptyset)$ \\
$Q(\mathrm{yes}:\emptyset) \gets R(a:\%u), R(b:\%u)$\`\qed
\end{tabbing}
\end{example}

\begin{example}[Unnesting] \label{exunnest}
Let $o$ be the object described by $R$.
The following single-rule program retrieves all
subobjects of $o$ (at arbitrary depths, but not $o$ itself)
that have a `name'-value
equal to `John'.  These objects are returned as top-level
elements of a result object $S$, with new keys generated by packing.
\begin{tabbing}
$S(\langle \$x \rangle.\$y:\%u) \gets
R(\$x.\mathrm{name}:\mathrm{John}), R(\$x.\$y:\%u)$
\end{tabbing}
\samepage
\qed
\end{example}

\begin{example}[Key lookup, nesting] \label{exnest}
Like the previous example,
the following program again considers subobjects,
but now focuses on those that have a key `ref' with an atomic key as
value.  That key is looked up and all values found for it are
collected in a new subobject created under the `ref' key.  As in
the previous example, new keys (for the elements of the
collection) are generated using packing.  The output object $S$
is thus an ``enrichment'' of the input object $R$.
\begin{tabbing}
$T(\$x.\mathrm{ref}:\emptyset) \gets
R(\$x.\mathrm{ref}:@k)$ \\
$S(\$x.\mathrm{ref}.\langle \$y \rangle.?z:\%u) \gets R(\$x.\mathrm{ref}:@k), R(\$y.@k.?z:\%u)$ \\
$S(\$x':\%u) \gets R(\$x':\%u), \neg T(\$x':\emptyset)$
\end{tabbing}
\end{example}

\section{Expressiveness and complexity} \label{secexpr}

Nonrecursive J-Logic has polynomial-time data complexity.  Since rules
are safe, we can find valuations satisfying the body of a rule
through finding valuations of predicates.  To find the valuations
satisfying a predicate $P(e:t)$, note that the path expression
$e$ is a sequence of key expressions and path variables.  Let $k$
be the length of this sequence.  Then we choose a pair $(p:v)$
from $P$; the number of possibilities is linear.  We match $t$ to
the atomic value $v$ in the obvious manner, and match $e$ to the
path $p$ by splitting $p$ in $k$ pieces.  The number of possible
splits is polynomial of degree $k$.  A piece corresponding to a
path variable provides a binding for that path variable, or must
be equal to an already existing binding.  A piece
corresponding to an atomic variable must be an atomic key.  A
piece corresponding to a constant must match the constant.
Finally, a piece corresponding to a packed key expression
$\langle e' \rangle$ must be a packed key $\langle p' \rangle$.
Then $e'$ is matched to $p'$ in turn.

Using positive, recursive, programs we can already simulate
Turing machines \cite{bonnermecca_sequences}.  Using general
programs, we are computationally complete: we can express any
computable $C$-generic query from finite instances to finite
instances.  Using an encoding of instances as defined here by
relational database instances, this can be proven following the known
body of work on the computational completeness of query languages
\cite{ch,av_proc,ak_iql,vvag_compl,cabibbo_stratified}.

\paragraph{Nonrecursive programs, relational algebra, and
practical languages}

Let us call a class of instances \emph{bounded} if there is a
fixed bound on the length of all paths occurring in the
instances, as well as on the nesting depth of packed keys.

On a bounded class of inputs, nonrecursive J-Logic can be
simulated by relational algebra.  Indeed, due to the bound, there
are only finitely many nonequivalent predicates, and each
equivalence class can be described using atomic variables only.
Thus, for each equivalence class of predicates we can keep the
bindings in a fixed-arity relation.  Given such a representation
the evaluation of a rule can be expressed in relational algebra.
Moreover, the application of a rule to a bounded instance
produces again a bounded instance (with the new bound depending
only on the old bound and the rule).  In this way we can simulate
nonrecursive J-Logic over bounded instances in relational
algebra.

Conversely, it is quite clear that we can represent all relational
database instances over some fixed schema as a bounded class of
instances in our data model. There are various ways
to do this.  One approach is to represent a tuple as an object in
the obvious way (each attribute is a key) and then represent a
set of tuples as a set of objects, using tuple identifiers as
top-level keys.  Under such a representation we can easily
simulate, say, the relational algebra, using nonrecursive
J-Logic.  We use packing to generate new tuple identifiers, as
illustrated in the Introduction for Cartesian product.

Another approach is to use a (bounded-depth) trie representation
for relations, as used, for example, in the Leapfrog Triejoin
algorithm \cite{leapfrogtriejoin}.  Such tries are naturally
represented as JSON objects.  We can then again simulate the
relational algebra using nonrecursive J-Logic, and we would not
even need packing.

Note that practical JSON query languages SQL++ \cite{sql++} and
JSONiq without recursive functions \cite{jsoniq_book} are mainly
geared towards bounded-depth data.  Apart from features such as
aggregation and full-text search, these languages are
fundamentally based on the nested relational algebra or calculus
\cite{bntw}.  This calculus can be translated into nonrecursive
J-Logic.  As already mentioned in the Introduction, packing can
be used to represent nested collections.  The only caveat (which
is also not really mentioned by SQL++ and JSONiq) is to do
duplicate elimination on nested collections.  It follows from
known results \cite{vdbp_abstr} that a special set-oriented
packing operator would need to be added for this purpose.

Moreover, we feel that the main contribution of J-Logic is as a
language in which nonrecursive programs can also work well with
unbounded inputs, i.e., deeply nested data.

\paragraph{Limitations of nonrecursive programs}

The above discussion immediately yields examples of queries not
expressible by nonrecursive programs: any query over relational
instances that is not expressible in the relational algebra will
do, such as the transitive closure of a binary relation.  That
does not tell us anything about unbounded instances, however.
In Proposition~\ref{theornec} we will give a general necessary
condition on the output of nonrecursive programs.

\begin{example} \label{exa}
Let $c$ be some constant and consider the query $Q$ from
$\{R\}$ to $\{S\}$ defined by
$$
Q(I) = \{S(k_1.c.k_2.c\dots k_n.c : \emptyset) \mid
R(k_1.k_2\dots k_n : \emptyset) \in I\}
$$
where $n$ is not fixed but ranges over all
possible lengths.  Proposion~\ref{theornec} will imply that this query
is not expressible by a nonrecursive program.\qed
\end{example}

Bonner and Mecca \cite{bonnermecca_sequences} have proposed
mixing transducers with sequence Datalog, so that manipulations
as in the above example can be easily expressed.
They already noted informally that without recursion through
concatenation, only a fixed number of concatenations can be
performed.  The following proposition formalizes this observation
and adapts it to J-Logic.

In order to state the necessary condition, we introduce the
following notations. For a set $S$ of paths, $\SP(S)$ denotes all
subpaths of paths occurring in $S$ (also paths occurring in
packed keys).  Also, $\CC Si$ denotes all paths that can be built
up (using concatenation and packing) from the paths in $S$ using
a total of at most $i$ concatenations.  The set of paths of an instance
$J$ is denoted by $\paths(J)$, so formally, $\paths(J) = \{p \mid
R(p:v) \in J$ for some $R$ and $v\}$.

\begin{proposition} \label{theornec}
Let $\Prog$ be a nonrecursive program.  There exists a finite set
$L$ of paths and a natural number $i$ such that for every
instance $I$, we have $\paths(\Prog(I)) \subseteq
\CC{\SP(\paths(I) \cup L)}i$.
\end{proposition}
\begin{proof}
By induction on the number of strata.  For the base case,
assume $\Prog$ consists of a
single stratum.  By an obvious rewriting we may assume without
loss of generality that the
body of each rule only mentions EDB relation names.  
Consider an element $p \in \paths(\Prog(I))$.  Then $p$
is produced by applying a
valuation to a path expression, say $e$, in the head of some
rule.  Every variable is mapped to an element of $\SP(\paths(I))$.
Let $\hat e$ denote the sequence obtained by removing all variables
from $e$, as well as all opening and closing brackets of packed
keys; we refer to these lexical elements as \emph{separators}.
Let $i_e$ denote the number of separators;
we can view $e$ as chopping $\hat e$ in $i_e+1$ pieces.
Thus, $p \in \CC{\SP(\paths(I) \cup \{\hat e\})}{i_e+1}$.  Hence,
we can set $i$ to the maximum $i_e$,
and we can set $L$ to the set of $\hat e$.

Now assume $\Prog$ has at least two strata.  Let $\Prog'$ be the
part without the last stratum, which we denote by $\Prog''$.  So,
$\Prog$ is the composition of $\Prog''$ after $\Prog'$.  By
induction, we have $i'$ and $L'$ for $\Prog'$.  Moreover,
reasoning as in the base case, we have $i''$ and $L''$ for
$\Prog''$ applied to $\Prog'(I)$.  After some calculations we can
see that we can now set $i=i'\cdot i''$
and $L = L' \cup L''$.
\end{proof}

\paragraph{Flat--flat queries}

An instance is called \emph{flat} if no packed keys occur in it.
A query $Q$ is called \emph{flat--flat} if for every flat
instance $I$, if $Q(I)$ is defined then it is also flat.
It may still be convenient to use packing in the computation of a
flat--flat query, as illustrated next.

\begin{example} \label{exff}
The query from Example~\ref{exa} is flat--flat.  Over flat
inputs, we can compute it by the following program:
\begin{tabbing}
$T(\langle @i \rangle . ?y:\emptyset) \gets R(@i.?y:\emptyset)$\\
$T(?x.@i.c.\langle @j \rangle.?y:\emptyset) \gets T(?x.\langle @i
\rangle.@j.?y:\emptyset)$ \\
$S(?x.@i.c:\emptyset) \gets T(?x.\langle @i \rangle:\emptyset)$
\end{tabbing}
We see that packing is conveniently used as a cursor to run through the
sequence.  With more effort, however, we can also compute the
query without using packing.   The trick is to use some constant
$a$ and to look for the longest consecutive sequence of $a$'s
occurring in any path in $R$.  Then a sequence of $a$'s one
longer than that can be used as a cursor.  The program is as
follows.
Since all predicates in the program will be of the form
$P(e:\emptyset)$, we abbreviate them as $P(e)$.
\begin{tabbing}
$\mathit{Sub}(a.?y) \gets R(?x.a.?y.?z)$ \\
$\mathit{Subnota}(\$x.@i.?y) \gets \mathit{Sub}(\$x.@i.?y),\
@i\neq a$ \\
$\mathit{Suba}(\$x) \gets \mathit{Sub}(\$x), \neg
\mathit{Subnota}(\$x)$ \\
$A(a.\$x) \gets \mathit{Suba}(\$x), \neg \mathit{Suba}(a.\$x)$ \\
$T(\$a.\$x) \gets R(\$x), A(\$a)$ \\
$T(?x.@i.c.\$a.?y) \gets T(?x.\$a.@i.?y), A(\$a)$ \\
$S(\$x) \gets T(\$x.\$a), A(\$a)$\`\qed
\end{tabbing}
\end{example}

The above example illustrates a general theorem:
\begin{theorem}[Flat--flat theorem] \label{theorff}
For every J-Logic program computing a flat--flat query there is
equivalent program without packing, over flat instances.
\end{theorem}

\begin{proof}
The proof is easy if we can use two constants, say $a$
and $b$, that are never used in any instance.  Then a packed key
expression $\langle e \rangle$ can be simulated using $a.e.b$,
where we also would need to write additional rules checking that $e$
matches a path with balanced $a$'s and $b$'s.  

If we want a simulation that always works, without an assumption
on the constants used in instances, we can
encode a path $p=k_1.k_2\dots k_n$ by
its doubled version $p'=k_1.k_1.k_2.k_2\dots k_n.k_n$.  Then
$\langle e \rangle$ can be simulated using $a.b.e'.b.a$.
\new{For example, the path $a.c.\langle a.b \rangle .b.a$ is
encoded as $a.a.c.c.a.b.a.a.b.b.b.a.b.b.a.a$.}
\new{This encoding} can be computed without packing using the technique
illustrated in Example~\ref{exff}.
\new{
Assume we want to encode the contents of a relation $A$ and
have in relation $A_c$ computed with this technique 
a path of $c$'s that is one longer
then the longest path in relation $A$. Let us call this path $c^{+1}$.
We can then define
a program, that computes the
encoding of $A$. This program starts with copying $A$ but
adds $d.c^{+1}.c^{+1}.d$ as a cursor with $d$ different from $c$,
and then moves this cursor
to the left while doubling constants. Note that we cannot use
$d.c^{+1}.d$ as a cursor since we are doubling paths and so
might be creating subpaths equal to $d.c^{+1}.d$. We
can also not simply use $c^{+1}.c^{+1}$ since the original
path might contain $c$'s and so there might be
uncertainty while matching about where the cursor begins and ends.
\begin{tabbing}
$\mathit{A_1}(\$x .d. c^{+1}.c^{+1}.d : \%u) \gets A(\$x : \%u)$ \\
$\mathit{A_1}(?x . d.c^{+1}.c^{+1}.d . @i . @i . ?y : \%u) \gets \mathit{A_1}(?x .@i . d.c^{+1}.c^{+1}.d . ?y : \%u)$
\end{tabbing}
Recall that we are encoding the input of a flat-flat query, and so
can assume the input contains no packing that needs to be
encoded. As a final step we then select those paths where
the cursor has arrived at the beginning and remove the cursor,
which produces the encoding
of $A$ in $A_2$.
\begin{tabbing}
$\mathit{A_2}(\$x : \%u) \gets A_1(d.c^{+1}.c^{+1}.d.\$x : \%u)$
\end{tabbing}

We can transform the original program with packing to one
that does not use packing and assumes that the input is encoded
as previously described. This transformation is done as follows:
\begin{itemize}
  \item Any constant and atomic variables in a path expression are
     doubled, like in the encoding.
    So a constant $a$ is replaced with $a.a$, and an atomic variable
    $@i$ is replaced with $@i.@i$. Note that in a predicate $P(e : t)$
    we do not replace constants and atomic
    variables in $t$. 
  \item Any path variable is left in place, but the clause is extended
    with a check $\mathit{Enc}_B(\$x)$ with $B$ a relation name of
    a predicate in which $\$x$ occurs in the rule, to see if the variable $\$x$
    matches a subpath of a path in $B$ that is a valid encoding.
  \item Any packed key expression $\langle e \rangle$ is replaced 
    with $a.b.e'.b.e$ where $e'$ is the transformation of $e$.
\end{itemize}
As an example, consider the following rule:
\begin{tabbing}
$\mathit{A}(@v.a.\$y : @w) \gets B(\$x.\langle c.\$y \rangle.@w : @v), 
  \neg C(@v.b : \emptyset)$
\end{tabbing}
It is translated to:
\begin{tabbing}
$\mathit{A}(@v.@v.a.a.\$y : @w) \gets B(\$x.a.b.c.\$y.b.a.@w.@w : @v), 
  \neg C(@v.@v.b.b : \emptyset), \mathit{Enc}_B(\$x), \mathit{Enc}_B(\$y)$
\end{tabbing}
The predicate $\textit{Enc}_B$ can be expressed by a program 
without packing as follows:
\begin{tabbing}
$\mathit{Enc}_B(a.b.?x.b.a) \gets B(?u.a.b.?x.b.a.?v), \mathit{Enc}_B(?x)$ \\
$\mathit{Enc}_B(@i.@i.?x) \gets B(?u.@i.@i.?x.?v), \mathit{Enc}_B(?x)$ \\
$\mathit{Enc}_B(?x.@i.@i) \gets B(?u.?x.@i.@i.?v), \mathit{Enc}_B(?x)$
\end{tabbing}
It is clear that the transformed program simulates
the original program on encoded instances.

As the following step we need to show that the encoded result can
be decoded without using packing. So let $B$ be a relation in
$\Vout$ that contains an encoded result,
and assume that with the technique of Example~\ref{exff}
we have computed in $B_c$ a path of $c$'s that is one longer then
the longest path of $c$'s in $B$. We will denote this path of $c$'s as
$c^{+1}$. The approach is basically the same as for the encoding: we
place a cursor in each path to indicate until how far we have decoded 
the path. The first program copies $B$ but
adds $d.c^{+1}.d$ as a cursor, and then moves this cursor
to the left while undoubling constants.
\begin{tabbing}
$\mathit{B_1}(\$x .d.c^{+1}.d : \%u) \gets B(\$x : \%u)$ \\
$\mathit{B_1}(?x .d.c^{+1}.d . @i . ?y : \%u) \gets \mathit{B_1}(?x .@i . @i . d.c^{+1}.d . ?y : \%u)$
\end{tabbing}
Recall that we are decoding the output of a flat-flat query, and so
can assume the input contains no encoded packing that needs to be
decoded. As a final step we select the paths where
the cursor has arrived at the beginning and remove the cursor,
which produces the decoding
of $B$ in $B_2$.
\begin{tabbing}
$\mathit{B_2}(\$x : \%u) \gets B_1(d.c^{+1}.d.\$x : \%u)$
\end{tabbing}
}
\end{proof}

The above proof needs recursion, even if the given program is
nonrecursive.  In general it is fair to say that the above flat--flat
theorem is mainly of theoretical interest.  Still
it is an interesting open question whether for every
nonrecursive program computing a flat--flat query, there is an
equivalent \emph{nonrecursive} program without packing, over all flat
instances.

\section{Proper object descriptions and object--object queries}
\label{secoo}

In Remark~\ref{remproper} we introduced the notion of proper
object description as the object description of an actual object,
as opposed to just any set of path--value pairs.
Proper object descriptions can be characterized as follows.

\begin{proposition} \label{proper}
A finite object description $D$ is proper if and only if
it satisfies the following two constraints:
\begin{itemize}
\item
the functional dependency
from paths to atomic values, i.e., if $(p:u)
\in D$ and $(p:v) \in D$, then $u=v$.
\item
prefix-freeness, i.e., if $p$ and $q$ are paths, and $(p.q:u)
\in D$ for some $u$, then $(p:v) \notin D$ for every $v$.
\end{itemize}
\end{proposition}

\begin{proof} The only-if
direction is clear.  The if-direction can be proven by induction
on the maximum length of a path in $D$.  If this maximum equals
$1$, then $D$ clearly describes an object with only atomic
values.  The object is well-defined thanks to the functional
dependency.  Now assume the maximum is at least $2$.  We construct an
object $o$ such that $\OD(D)=o$ as follows.

Define $K_1$ as the
set of keys $k$ such that
$(k:v) \in D$ for some $v$.  Thanks to the functional dependency
$v$ is unique for $k$ and we denote $v$ by $D(k)$.  As in the
base case, we obtain an object $o_1$ defined on $K_1$ defined by
$o_1.k=D(k)$.

Define $K_2$ as the set of atomic keys $k$ such that 
$(k.p:v) \in D$ for some path $p$ and atomic value $v$.
For each $k \in K_2$ define the object description $D_k = \{(p:v)
\mid (k.p:v) \in D\}$.  Then $D_k$ has a shorter maximum path
length and still satisfies the two constraints.  Hence, by
induction, $D_k$ describes an object $o_k$.  We now define the
object $o_2$, defined on $K_2$, by setting $o_2.k = o_k$.

Thanks to prefix-freeness, $K_1$ and $K_2$ are disjoint.
Hence the union $o_1 \cup o_2$ is a well-defined object and
yields the desired object $o$.
\end{proof}

\begin{example}
$D = \{a:1,a.a:1\}$ is not prefix-free and
indeed $D$ is not proper.  In proof,
suppose $D$ would be the description of an object $o$.
Then $o.a$ is the atomic value $1$ by the first pair in $D$.
But by the second
pair, $o.a$ is an object with $a$-value 1, a contradiction.\qed
\end{example}

\begin{remark}
Only finite object descriptions can be proper, since objects are
always finite.  Still, the
two constraints from the above proposition can be taken to be
the \emph{definition} of properness for infinite instances.
Later in this paper, we will consider the
object--object problem, the implication problem for jaegds,
and the containment problem.  These three problems
ask a question about \emph{all} instances.  These
problems do not change, however, if we restrict
attention to finite instances.\qed
\end{remark}

An instance is called proper if it assigns a proper object
description to every relation name. 
A query $Q$ is called
\emph{object--object} if for every proper instance $I$, if 
$Q(I)$ is defined then it is also proper.

The object--object property is practically important.  In
practice, a JSON processor may accept improper object
descriptions, or object syntax that is not well-defined, such as
$\{a:1,a:2\}$ or $\{a:1,a:\{b:2\}\}$.  However, the processor
will interpret such syntax in an unpredictable manner.  Perhaps
it will overwrite a previously read $a$-value by an $a$-value
read later.  Or, on the contrary, it may keep only the value that
was read first.  To avoid depending on such system-defined
behavior, we better write queries having the
object--object property.

One may go further and demand that also all intermediate
relations generated by a J-Logic program hold proper object
descriptions.  This may be relevant, for example, if we implement
the query language on top of a JSON store.  We next show that
this is always possible.  We call the result
the ``object--object theorem'', which may be a bit pompous, as
it is proven by a simple trick using packing (thus again
illustrating the utility of packing).

\begin{theorem}[Object--object theorem] Let $\Prog$ be a program
expressing an object--object query $Q$.  Then there exists an
equivalent program $\Prog'$ such that,
on any proper input instance, all IDB relations 
of $\Prog'$ hold proper object descriptions. Program $\Prog'$ has the
same number of strata as $\Prog$, and is recursive only if
$\Prog$ is.
\end{theorem}
\begin{proof}
The idea is to encode arbitrary object descriptions by
object descriptions that are always proper.
Then the program is simulated using the encoding.
At the end the output relations are decoded.
Such an encoding is easy to do using packing.

Formally, fix an arbitrary
atomic key $b$.  For any input
relation name $R$ we introduce the following two encoding rules:
\begin{tabbing}
$R'(\langle \$x \rangle.\langle @u \rangle:\emptyset) \gets R(\$x:@u)$ \\
$R'(\langle \$x \rangle.\langle b.b \rangle:\emptyset) \gets
R(\$x:\emptyset)$
\end{tabbing}
These rules are added to the first stratum of $\Prog$.

Furthermore, we modify $\Prog$ by replacing each atom (in bodies
and in heads) of the form
$P(e:t)$ by $P'(\langle e \rangle.\langle t \rangle:\emptyset)$ if
$t$ is not $\emptyset$, and by $P'(\langle e \rangle.\langle
b.b\rangle:\emptyset)$ otherwise.

Finally for every output relation name $S$ we add the following
decoding rules to the last stratum:

\medskip \noindent
$S(\$x:@u) \gets S'(\langle \$x \rangle.\langle @u \rangle)$ \\
$S(\$x:\emptyset) \gets S'(\langle \$x \rangle.\langle b.b
\rangle)$\qedhere
\end{proof}

\begin{example}
The following program begins by eliminating the top layer from an
object $R$, which brings the second-level keys to the top level.
This intermediate result $R_1$ may well be improper. We then
throw away all ``bad'' paths (paths that violate properness).
The result, $S$, is of course proper.  Thus, this program
computes an object--object query but is easiest to write using
improper intermediate results.  Yet, the object--object theorem
assures us it can be rewritten using only proper intermediate
results.
\begin{tabbing}
$R_1(\$y : \%u) \gets R(\#x.\$y:\%u)$ \\
$\mathit{Bad}(\$y:\%u) \gets R_1(\$y:\%u), R_1(\$y:\%v),
\%u\neq\%v$ \\
$\mathit{Bad}(\$x.\$z:\%v) \gets R_1(\$x:\%u), R_1(\$x.\$z:\%v)$
\\
$S(\$y:\%u) \gets R_1(\$y:\%u), \neg \mathit{Bad}(\$y:\%u)$\`\qed
\end{tabbing}
\end{example}

\begin{remark}

Our proof of the object--object theorem uses packing.  Of course
there is nothing wrong with packing; we think it is a versatile
tool.  Yet, theoretically one may wonder whether one can also do
without.  Indeed it turns out one can prove a combination of the
flat--flat theorem and the object--object theorem.  Specifically,
for every program computing a flat--flat object--object query, we
can find a program without packing that is equivalent over flat
instances and that only works with proper intermediate results.
The idea is to encode a path--value pair $a_1\dots a_n:\emptyset$
by $b.a_1\dots b.a_n.a.a.b:\emptyset$, and a path--value pair
$a_1\dots a_n:c$ by $b.a_1\dots b.a_n.c.a.a.b:\emptyset$.  It can
be verified that an encoding of an object description is always
proper.  The program is then modified to work over encodings.
As for the flat--flat theorem, the program without packing would
need recursion.  Again we leave open whether there is a
nonrecursive version of the flat--flat object--object theorem.

\end{remark}

\subsection{Deciding the object--object property} \label{sect:object-object}

The \emph{object--object problem} is to decide,
given a J-Logic program $\Prog$ and appropriate
vocabularies $\Vin$ and $\Vout$,  whether the query from $\Vin$
to $\Vout$ computed by $\Prog$ has the object--object property.

In general, this problem is of course undecidable. It is
undecidable for positive recursive programs, because these can
simulate Turing machines, and also for nonrecursive programs that can
use negation, because these can express first-order logic
(relational algebra).

\new{
Another restriction we will introduce concerns the use of equations in
programs. These can sometimes add expressive power
that is usually associated with recursive programs.
\begin{example} \label{example:cyclic1}
The following program selects from $R$ all paths that contain
only $a$'s.
\begin{tabbing}
$S(\$x : \%u) \gets R(\$x : \%u), a.\$x = \$x.a$ \qed
\end{tabbing}
\end{example}
To rule out such programs we introduced the following definitions.
Given a rule $H \gets B$ we define the \emph{equation graph} as
an undirected multigraph where all variables in $B$ are the nodes
and the number of edges between variable $x$ and variable $y$ is
equal to the sum of $\#_x(e_1) \times \#_y(e_2)$ for each distinct equation 
$e_1 = e_2$ in $B$, where $\#_x(e)$ denotes the number of times
variable $x$ occurs in $e$. We call a nonempty sequence of edges in an
equation graph a \emph{path} if in the sequence each two subsequent
edges are incident. We call
a path a \emph{cycle} if the first and last edge are incident, and
an equation graph \emph{cyclic} if it contains a cycle. 
We say that the rule $H \gets B$ is \emph{equationally cyclic} if
the equation graph associated with $B$ contains a cycle. We call
a program \emph{equationally cyclic} if at least one of its rules is cyclic,
and \emph{equationally acyclic} if there is no such rule.
\begin{example}
The following rule is equationally cyclic:
\begin{tabbing}
$S(\$x : \%u) \gets R(\$x : \%u), a.\$x = \$y, \$y = \$x.a$
\end{tabbing}
This is because its equation graph contains
the cycle $\langle \{ \$x, \$y \}_{a.\underline{\$x} = \underline{\$y}}, 
\{ \$y, \$x \}_{\underline{\$y} = a.\underline{\$x}} \rangle$. The
subscript of each edge indicates which equation and
occurrences the edge corresponds to.

Also the following rule has a cyclic equation graph:
\begin{tabbing}
$S(\$x : \%u) \gets R(\$x : \%u), \$y.\$y = \$x$
\end{tabbing}
This is because it contains the cycle 
$\langle \{ \$x, \$y \}_{\underline{\$y}.\$y = \underline{\$x}}, 
\{ \$y, \$x \}_{\$y.\underline{\$y} = \underline{\$x}} \rangle$.

The program in Example~\ref{example:cyclic1} is also equationally cyclic since
the equation graph of its rule contains the cycle 
$\langle \{ \$x \}_{a.\underline{\$x} = \underline{\$x}.a} \rangle$.  \qed
\end{example}

This restriction on program allows us to formulate the main result
of this subsection:
}
\begin{theorem} \label{oodecide}
The object--object problem is decidable for
positive, nonrecursive programs \new{where all rules are equationally
acyclic and their head contains every variable at most once}.
\end{theorem}
Our starting point is to note
that this problem has similarities with a problem known from
relational databases.  This problem is the FD--FD implication
problem for (unions of) conjunctive queries (UCQs)
\cite{ah_functions,ahv_book}. It is also called the view dependency
problem \cite{klugprice}.  This problem asks, given two sets
$\Sigma_1$ and $\Sigma_2$ of functional dependencies (FDs) and a
query $Q$, whether the result of $Q$, applied to an instance
satisfying $\Sigma_1$, always satisfies $\Sigma_2$.  The
similarity lies in that properness involves
satisfying an FD; moreover, positive nonrecursive J-Logic
programs \new{that are equationally acyclic} are the J-Logic analog of UCQs.
Of course there are also differences: J-Logic has packing and
path variables, and the notion of properness is not only about
FDs but also about prefix-freeness.

The decidability of the FD--FD implication problem for UCQs
follows readily from the decidability of the implication problem for
equality-generating dependencies (egds), using the chase
\cite{ahv_book,bv_proof_procedure}.  Hence our approach is to
introduce \emph{J-Logic atomic
equality-generating dependencies} or jaegds, and investigate
the chase for these dependencies.

Syntactically, a \emph{jaegd} is a rule $\sigma$ of the form $B
\to E$, where $B$ is a positive body without equalities and $E$
is an atomic equality, i.e., an equality of the form $u=v$ where
$u$ and $v$ are atomic key expressions
(atomic constants or atomic variables).  If $u$ or $v$
is a variable, that variable must occur in $B$.

Semantically, note that $B$ consists exclusively of positive
predicates.  Hence, for any instance $I$ and valuation $\nu$
appropriate for $B$, we have that $I,\nu$ satisfies $B$ if and
only if $\nu(B) \subseteq I$.  We denote this by $\nu : B \to I$
and call $\nu$ a \emph{matching} of $B$ in $I$.  We now define
that $I$ satisfies a jaegd $\sigma$ as above,
denoted by $I \models \sigma$, if
for every matching $\nu : B \to I$,
the atomic keys $\nu(u)$ and $\nu(v)$ are identical.

Note that dependencies of the form $B \to a=b$,
where $a$ and $b$ are distinct atomic keys, are allowed.  Since $a=b$ is
always false, this can be written more clearly as $B \to
\mathbf{false}$ or also $B \to \bot$.  This is used to
express a \emph{denial constraint}: it is only satisfied in an
instance $I$ if there does not exist any matching of $B$ in $I$.

Note that we also allow dependencies of the form $B \to u=u$.
Obviously such dependencies are
trivial (satisfied in any instance), but we allow them because
they may be produced by the chase procedure.

\begin{example} \label{exdelta}
By Proposition~\ref{proper}, an 
object description $D$ is proper if and only if it satisfies
the jaegds $\delta_1$--$\delta_6$:
\begin{itemize}
\item[]
$\delta_1 : D(\$x:@i), D(\$x:@j) \to @i=@j$ \\
$\delta_2 : D(\$x:\emptyset) , D(\$x:@i) \to \bot$ \\
$\delta_3 : D(\$x:@i) , D(\$x.\$y:\emptyset) \to \bot$ \\
$\delta_4 : D(\$x:@i) , D(\$x.\$y:@j) \to \bot$ \\
$\delta_5 : D(\$x:\emptyset) , D(\$x.\$y:\emptyset) \to \bot$ \\
$\delta_6 : D(\$x:\emptyset) , D(\$x.\$y:@j) \to \bot$\qed
\end{itemize}
\end{example}

For a set of dependencies $\Sigma$, we define $I \models \Sigma$
to mean that $I$ satisfies every dependency in $\Sigma$.  We say
that $\Sigma$ \emph{logically implies} a dependency $\sigma$ if
every instance that satisfies $\Sigma$ also satisfies $\sigma$.
The \emph{implication problem for jaegds} asks to decide, given a
set of jaegds $\Sigma$ and a jaegd $\sigma$, whether $\Sigma$
logically implies $\sigma$.  We actually do not know whether this
problem is decidable in general.  We will, however, solve a
special case that is sufficient to solve the object--object
problem.

\paragraph{The Chase}
We first
need the notion of a \emph{variable mapping}.  This is a function
defined on a finite set of variables that maps path variables to
path expressions and atomic variables to atomic key
expressions.  Like valuations, we can apply a variable mapping to a
predicate simply by applying it to every variable occurring in
the predicate.  The result is again a predicate. Thus, the result
of applying a variable mapping to a body is again a body.
A \emph{homomorphism} $h$ from a body $B_1$ in a body $B_2$,
denoted by $h:B_1\to B_2$, is a variable mapping defined on at
least all variables in $B_1$ such that $h(B_1) \subseteq B_2$.

With this notion of homomorphism in place, the notion of
\emph{chasing a jaegd $\sigma$ with a set of jaegds $\Sigma$} is
defined entirely similarly to the well-known chase for egds in the
relational model \cite{ahv_book}.  

\paragraph{The Chase}

Let $\Sigma$ be a set of jaegds and
let $\sigma$ be a single jaegd.  Let $B$ be the body of $\sigma$.
By \emph{applying a chase step} we mean the following:
\begin{enumerate}
\item
Pick a dependency $C \to (u=v)$ in $\Sigma$.
\item
Pick a homomorphism $h : C \to B$ such that $h(u)$ and $h(v)$ are
not identical.
\item
We consider the possibilities:
\begin{itemize}
\item
If $h(u)$ and $h(v)$ are different atomic keys, we say that the 
chase step has failed.
\item
If one of $h(u)$ and $h(v)$ is an atomic key and the other is a variable, we
substitute the atomic key for the variable everywhere in $\sigma$.
\item
If both $h(u)$ and $h(v)$ are variables, we substitute $h(u)$ for
$h(v)$ everywhere in $\sigma$.
\end{itemize}
\end{enumerate}

If we can apply a sequence of chase steps, starting in $\sigma$,
and applying each subsequent step to the result of the previous
step, until we can make the chase step fail, we say that \emph{chasing
$\sigma$ with $\Sigma$ fails}.
If, in contrast, we can apply a sequence of chase steps without
failure until no chase step can be applied anymore, we say that
chasing $\sigma$ with $\Sigma$ succeeds.  An infinite sequence of
chase steps is not possible, because we only equate atomic
variables to atomic keys and the number of atomic variables and
keys appearing in $B$ is finite.

It is not difficult to see that the chase is locally confluent,
whence confluent by Newman's Lemma.  Hence, given the above
definitions, it is not possible for the chase to succeed and fail at
the same time.

The chase provides a sound proof
procedure for logical implication, as stated in the following
proposition.  

\begin{proposition} \label{propchaseif}
Assume that either chasing $\sigma$ with $\Sigma$ fails, or
the chase succeeds and results in a jaegd whose consequent is a
trivial equality.  Then $\Sigma$ logically implies $\sigma$.
\end{proposition}

The proof for this proposition is essentially the same as 
for egds in the relational model. It starts with the following
property, which expresses soundness of the chase procedure.

\begin{lemma} \label{lemsoundchase}
If the chase fails, then $\sigma$ is vacuously true under
$\Sigma$, i.e., for every instance $I$ satisfying $\Sigma$, there
exists no matching of $B$ in $I$.
If the chase succeeds with a final result $\sigma'$,
then $\sigma$ and $\sigma'$ are equivalent under $\Sigma$, i.e.,
for every instance $I$ satisfying $\Sigma$, we have $I \models
\sigma$ if and only iff $I \models \sigma'$.
\end{lemma}

The above lemma implies the following:

\begin{proof}[Proof of Proposition~\ref{propchaseif}]
Let $I$ be an instance satisfying $\Sigma$.
We must show $I \models \sigma$.  Thereto consider a matching
$\alpha : B \to I$.  By Lemma~\ref{lemsoundchase}, chasing $\sigma$ 
by $\Sigma$ succeeds (otherwise the matching $\alpha$ cannot
exist).  We are given that the chase yields a dependency
$\sigma'$ with a trivial equality as a consequent.
Hence, trivially $I \models \sigma'$.  However, by
Lemma~\ref{lemsoundchase}, this implies also $I \models \sigma$
as desired.
\end{proof}

For egds in the relational model, the converse to the above
proposition holds as well, showing the completeness of the chase
as a proof procedure.  In our model, however, the converse fails,
as shown next.

\begin{example} \label{exfail}
Consider $\Sigma$ consisting of the following three denial
constraints:
\begin{align*}
& P(@x:\emptyset) \to \bot \\
& P(\langle \$x \rangle:\emptyset) \to \bot \\
& P(\$x.\$y :\emptyset) \to \bot
\end{align*}
Then $\Sigma$ is equivalent to the single denial constraint
$\sigma \equiv P(\$x:\emptyset) \to \bot$, so certainly $\Sigma$
logically implies $\sigma$.  However, chasing
$\sigma$ with $\Sigma$ does not fail. Actually, no chase step can
be applied at all and the chase ends immediately on $\sigma$
itself.  Since the consequent $\bot$ is not a trivial equality,
this shows that the converse of Proposition~\ref{propchaseif}
fails.\qed
\end{example}

We can still get completeness of the chase in a special case,
which we call \emph{unambiguous}.  We first define the
notion of \emph{weak variable mapping}.  Recall that a variable
mapping must map atomic variables to atomic key expressions.
A weak variable mapping
is like a variable mapping, except that atomic variables may also
be mapped to path variables.  A \emph{weak morphism} from a body $B_1$ in a
body $B_2$ is a weak variable mapping $h$ such that $h(B_1)
\subseteq B_2$.

Now consider an input $(\Sigma,\sigma)$ to the implication
problem for jaegds.  We say that $(\Sigma,\sigma)$ is
\emph{unambiguous}
if either chasing $\sigma$ with $\Sigma$ fails, or the chase
succeeds, and the following condition holds.  Let $B'$ be the
body of the jaegd resulting from the chase.  Then every weak
morphism from a body in $\Sigma$ to $B'$ must actually be a variable
mapping.

\begin{example}
Take $\Sigma$ and $\sigma$ from
the previous example.  We already noted that the chase succeeds
immediately.
We see there is a weak morphism
from the body $\{P(@x:\emptyset)\}$ of the first dependency in
$\Sigma$, to the body $\{P(\$x:\emptyset)\}$ of $\sigma$, namely the
mapping $@x \mapsto \$x$.  This is not a variable mapping.
Hence $(\Sigma,\sigma)$ is not unambiguous.\qed
\end{example}

\begin{example} \label{exdeltacomplete}
For another example, consider the set
$\Delta=\{\delta_1,\dots,\delta_6\}$ from Example~\ref{exdelta}.
Then $(\Delta,\sigma)$ is always unambiguous for any $\sigma$.
Indeed,
atomic variables occur only in the second component of
predicates in $\Delta$, i.e., after the $:$ sign, and path
variables can never occur after the $:$ sign in any body.\qed
\end{example}

The notion of unambiguity captures the cases where the usual
proof of completeness of the chase applies in our setting.  So, we
have the following result.  

\begin{proposition} \label{propchasecomplete}
Assume $\Sigma$ logically implies $\sigma$, and $(\Sigma,\sigma)$
is unambiguous.  Then chasing $\sigma$ with $\Sigma$ fails, or
the chase succeeds and results in a jaegd whose consequent is a
trivial equality.
\end{proposition}

\begin{proof}[Proof of Proposition~\ref{propchasecomplete}]
Let $\sigma$ be of the form $B \to (w=z)$.
Suppose the chase succeeds and results
in the jaegd $\sigma' \equiv B' \to (w'=z')$.
We must prove that $w'$ and $z'$ are identical.

We can view $B'$ as an instance $I$ by viewing each variable as
an atomic key; it is customary to refer to these atomic keys
as \emph{frozen variables}.  We claim that $I \models \Sigma$.

To prove the claim, consider a
dependency $C \to (u=v)$ in $\Sigma$ and a matching
$\alpha : C \to I$.  We can view $\alpha$ as a weak morphism from
$C$ to $B'$.  Because $(\Sigma,\sigma)$ is unambiguous, $\alpha$
does not map atomic variable to frozen path variables, i.e., it
is really a homomorphism from $C$ to $B'$.  Since the chase
succeeded with $B'$ the body of the final result, there is no
chase step possible in $B'$.  This means that $\alpha(u)$ and
$\alpha(v)$ must be identical and thus $I \models \phi$.

We now know that $I \models \Sigma$.  Since we are given that $\Sigma$
logically implies $\sigma$, also $I \models \sigma$.  Recall that
$\sigma'$ is the result of subsequent applications of chase
steps, starting from $\sigma$.  Each chase step maps an atomic
variable to another atomic variable or an atomic key, so amounts
to applying a homomorphism.  The composition of homomorphisms is
also a homomorphism.  Hence, there is a homomorphism from $B$ to
$B'$ that maps $w$ to $w'$ and $z$ to $z'$.  This homomorphism
can be viewed as a matching of $B$ in $I$.  Since $I \models
\sigma$, the images of $w$ and $z$ must be identical.  We
conclude that $w'$ and $z'$ are identical as desired.
\end{proof}

It follows that
the unambiguous cases of the implication problem for jaegds are
decidable by the chase.  En route to solving the object--object
problem, it is especially important that chasing from $\Delta$ is
unambiguous, as we saw in Example~\ref{exdeltacomplete}.

\paragraph{Equality elimination}
There is one final hurdle to overcome.
A discrepancy between bodies of jaegds and bodies of positive J-Logic
rules that are equationally acyclic is that the latter can have equalities.
We next show,
however, that equalities can always be removed.

\new{
Consider the equality $e_1 = e_2$ where $e_1 = \$x.@y.a.b.\$x$ and $e_2
 = \$v.@w.b.\$u$.  We define a notion of \emph{unifier} as a variable mapping
 that, when applied as a substitution, maps two path expressions to the same 
 path expression. For example,
 for $e_1$ and $e_2$ we have the following unifier: 
 $u_1 = \{ \$v \mapsto \$x.@y, @w \mapsto a, \$x \mapsto \$u \}$. Note that
 indeed $u_1(e_1) = \$x.@y.a.b.\$u = u_1(e_2)$.  We say that a unifier $u_1$ is
 equal or more general than another unifier $u_2$ if there is a variable 
 mapping $u_3$ such that $u_2(e) = u_3(u_1(e))$ for any
 path expression $e$. 
 We call two unifiers equivalent if one is equal or more general than the other
 and vice versa.  It is not hard to see
 that this defines a pre-order and moreover that if two unifiers are equivalent,
 they must be identical up to renaming the variables in the result.
 We will call a unifier a \emph{most-general unifier} if all unifiers that are
 equal or more general are in fact equally general.
 
Then, we can observe the following:
 
 \begin{lemma} \label{lem:finunif}
Given an acyclic equality $e_1 = e_2$ where $e_1$ and $e_2$ 
then the set of most-general unifiers of $e_1$ and $e_2$ has finitely many
equivalence classes.
\end{lemma}

\begin{proof}
We start with considering a unifier of $e_1$ and $e_2$ that maps them
both to a path expression $e_3$. For example,
let us consider $e_1 = a.\$x . b . \$y . c$ and 
$e_2 = a.\langle \$u \rangle . \$v . \langle @w \rangle . b. c$. A possible
unifier $u$ that maps both to a path expression $e_3$ can be represented
in a diagam as follows:

\begin{center}
\bgroup
\setlength\tabcolsep{0.6pt}
\begin{tabular}{ccccccccccccccccccccccccccccccccccc}
& & & \multicolumn{9}{c}{\scriptsize $\$x$} & & & &  \multicolumn{13}{c}{\scriptsize $\$y$} \\[-5pt]
& & & \multicolumn{9}{c}{\downbracefill} & & & &  \multicolumn{13}{c}{\downbracefill} \\
$e_3 =$ &$a$ & $.$ & $\langle$ & $b$ & $.$ & $@d$ & $\rangle$ & $.$ & $a$ & $.$ & $\$e$ & $.$ & $b$ & $.$ & 
$\langle$ & $b$ & $.$ & $\$e$ & $\rangle$ & $.$ & $@d$ & $.$ & $\langle$ & $@d$ & $\rangle$ & $.$ & $b$ & $.$ & $c$ \\[-5pt]
& & & & \multicolumn{3}{c}{\upbracefill} & & & \multicolumn{13}{c}{\upbracefill} & & &  \multicolumn{1}{c}{\upbracefill} & \\[-2pt]
& & & & \multicolumn{3}{c}{\scriptsize $\$u$} & & & \multicolumn{13}{c}{\scriptsize $\$v$} & & & \multicolumn{1}{c}{\scriptsize $@w$} &
\end{tabular}
\egroup
\end{center}

We can observe that a fragment of $p$ where two variables overlap,
such as for example the fragment $\langle b . \$e \rangle . @d$ where
$\$y$ and $\$v$ overlap, it holds that this fragment is well-balanced.
This is because every opening bracket in the fragment must
have a following matching closing bracket in the fragment, since the
fragment of $\$v$ is well-balanced. Vice versa, every closing bracket
in the fragment must have preceding matching opening bracket in the
fragment, since the fragmetn of $\$y$ is well-balanced. Consequently,
the fragment in the overlap is a path. 

It follows that from the unifiers we can derive a more general unifier $u'$
by replacing in the diagram every fragment where two variables overlap with a
distinct fresh variable. This fresh is a path variable, unless 
one of the two overlapping variables is an atomic variable, in which case
it is an atomic variable. In the previous example, 
this results in:

\begin{center}
\bgroup
\setlength\tabcolsep{0.6pt}
\begin{tabular}{ccccccccccccccccccccccccccccccccccc}
& & & \multicolumn{5}{c}{\scriptsize $\$x$} & & & & \multicolumn{7}{c}{\scriptsize $\$y$} \\[-5pt]
& & & \multicolumn{5}{c}{\downbracefill} & & & & \multicolumn{7}{c}{\downbracefill} \\
$e_4 =$ &$a$ & $.$ & $\langle$ & $\$q$ & $\rangle$ & $.$ & $\$r$ & $.$ & $b$ & $.$ & $\$s$ & $.$ & $\langle$ & $@t$ & $\rangle$ & $.$ & $b$ & $.$ & $c$ \\[-5pt]
& & & & \multicolumn{1}{c}{\upbracefill} & & & \multicolumn{5}{c}{\upbracefill} & & &  \multicolumn{1}{c}{\upbracefill} & \\[-2pt]
& & & & \multicolumn{1}{c}{\scriptsize $\$u$} & & & \multicolumn{5}{c}{\scriptsize $\$v$} & & & \multicolumn{1}{c}{\scriptsize $@w$} &
\end{tabular}
\egroup
\end{center}

It can be shown that the resulting diagram defines a unifier
if $e_1 = e_2$ is acyclic.
After all, if for two overlapping variables we make a replacement,
it follows from acyclicity that each of these variables occurs at most once
in $e_1$ and $e_2$. So there is only one place in $p'$ that describes
what these variables are mapped to, and so it is well defined what
they are mapped to after the replacement.

It will also be clear that this unifier $u'$  will be equally or or more general
than the original unifier $u$, since we obtain $e_3$ again if
we follow it with the substitution that replaced each new variable
with the fragment it replaced.

The number of equivalence classes of unifiers that are
generated by the previous
process can be shown to be finite. To show this, we introduce the concept
of \emph{symbol ordering}. By this we mean a linear order over the keys
and bracket occurrences in $e_1$ and $e_2$ that (1) allows occurrences
from $e_1$ to be merged with occurrences from $e_2$ if they concern
the same symbol and (2) respects the linear order of the occurrences
in $e_1$ and $e_2$. As an example of a symbol ordering consider the
following ordering, where the central horizontal line indicates the
linear order. Here the solid lines indicate the ordering defined by
$e_1$ and $e_2$ in the previous example, and the dahsed edges
indicated the added orderings to make it linear.

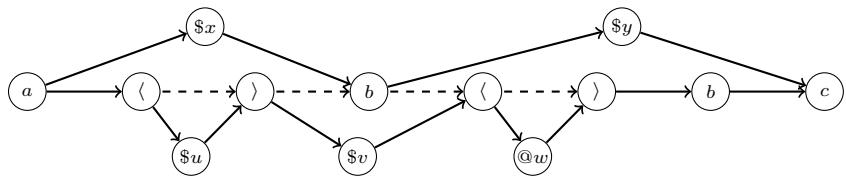
\begin{center}
\begin{tikzpicture}[
  every node/.style={circle, draw, minimum size=0.5cm, inner sep=0}, font=\scriptsize]
]
\node (a1) [circle, draw] {$a$};
\node (la1) [circle, draw, right=of a1] {$\langle$}
  edge [<-, thick] (a1);
\node (ra1) [circle, draw, right=of la1] {$\rangle$}
  edge [<-, dashed, thick] (la1);
\node (u) [circle, draw, below right=0.5cm and 0.3cm of la1]{$\$u$}
  edge [<-, thick] (la1) edge [->, thick] (ra1);  
\node (b1) [circle, draw, right=of ra1] {$b$}
  edge [<-, thick, dashed] (ra1);
\node (x) [circle, draw, above right=0.5cm and 2cm of a1]{$\$x$}
  edge [<-, thick] (a1) edge [->, thick] (b1);
\node (la2) [circle, draw, right=of b1] {$\langle$}
  edge [<-, thick, dashed] (b1);
\node (ra2) [circle, draw, right=of la2] {$\rangle$}
  edge [<-, thick, dashed] (la2);
\node (b2) [circle, draw, right=of ra2] {$b$}
  edge [<-, thick] (ra2);
\node (c1) [circle, draw, right=of b2] {$c$}
  edge [<-, thick] (b2);
\node (y) [circle, draw, above right=0.5cm and 3cm of b1]{$\$y$}
  edge [<-, thick] (b1) edge [->, thick] (c1);  
\node (v) [circle, draw, below right=0.5cm and 1cm of ra1]{$\$v$}
  edge [<-, thick] (ra1) edge [->, thick] (la2);  
\node (w) [circle, draw, below right=0.5cm and 0.3cm of la2]{$@w$}
  edge [<-, thick] (la2) edge [->, thick] (ra2);  
\end{tikzpicture}
\end{center}

It is easy to see that every generated diagram for $e_1 = e_2$ will
define a symbol ordering in its central horizontal line. 
Moreover, the symbol ordering, along
with the original linear order within $e_1$ and $e_2$,
completely determines the diagram since all that is required is
to select fresh variables for the dashed edges. This implies
that this also determines then the
unifier it defines. Since the linear orders in $e_1$ and $e_2$
can only be combined into a symbol ordering in finitely many
ways, it follows that there are only a finite number of
distinct (up to the choice of the fresh identifiers) unifiers
that are generated by the described process for generalising
unifiers.
\end{proof}
}

The previous result allows us to show that we can remove
equations from sets of equationally acyclic  rules.

\begin{lemma} \label{eqelim}
Every J-Logic rule \new{that is equationally acyclic} is equivalent to 
a finite set of equality-free
rules.  Also, every jaegd where we would allow equalities in
the body \new{such that it is equationally acyclic}, 
is equivalent to a finite set of equality-free jaegds.
\end{lemma}

\begin{proof}
\new{
We show by induction that a equationally acyclic rule with $n > 0$ 
equations, can be rewritten to an equivalent set of equationally
acyclic rules with $n - 1$ equations.

Let us consider a rule with equation $e_1 = e_2$. By 
Lemma~\ref{lem:finunif} we know that there is a finite
set of equivalence classes of most-general unifiers of
$e_1$ and  $e_2$ . We can select for each equivalence
class a representative that maps variables to path
expression with only fresh variables.

From the initial rule with the equation $e_1 = e_2$ 
we generate now a set of rules by (1) removing this
equality and (2) generate a rule for each unifier in the 
set of unifiers by applying it to the remainder of
the rule. Recall that a valuation, a function that maps
atomic variables to atomic keys and path variables to
paths, satisfies $e_1 = e_2$ iff it maps $e_1$ and $e_2$
to the same path. It follows that this holds iff the valuation
is a unifier of $e_1$ and $e_2$, which in turn holds iff
the valuation is an equal or less general unifier than
one of the most-general unifiers. It follows that replacing
the initial rule with the generated set of rules does
not change the semantics.

As a final step we show that the resulting set of rules
remains equationally acyclic. Let us consider one of
the newly generated rules, and assume it was generated
by the unifier $u$. Assume that the application of $u$
caused a cycle in the equation graph of the generated
rule. We can them map this cycle back to a cycle
that existed in the equation graph of the initial rule
as follows:
\begin{itemize}
  \item Consider an edge between two variables caused by
    occurrences
    that already existed before the application of $u$. Then
    the corresponding edge already existed in the equation
    graph of the initial rule.
  \item Consider an edge between an old variable $x$ and a
    new variable $y$ added by $u$. Let $z$
    be the unique variable
    that was replaced with a path expression containing
    $y$. Then, an edge between $z$ and $x$
    already existed in the equation graph of the initial rule.
  \item Consider an edge between new variable $x$ and
    new variable $y$. Let $v$ and $w$ be the unique
    variables that were replaced to introduce $x$ and $y$,
    respectively. If $v$ and $w$ are on opposite sides in
    $e_1 = e_2$ then there is a corresponding edge 
    between $v$ and $w$ in the old equation graph. If $v$ 
    and $w$ are on the same side, then there must be
    a variable $v'$ with which $v$ overlapped to generate
    $x$ and which is on the other side than $v$. It follows
    that there is an edge between $v$ and $v'$, and
    between $v'$ and  $w$ in the old equation graph.
\end{itemize}
In all considered cases it holds that for every edge
in the new equation graph there is a corresponding
edge or path in the old equation graph if we map
new variables back to the old variable that generated
them. It follows that for every cycle in the new
equation graph there must have already been a
corresponding cycle in the old equation graph.
}
\end{proof}

We are now ready for the
\begin{proof}[Proof of Theorem~\ref{oodecide}]
Let $\Prog$ be \new{an equationally acyclic} program 
computing a query $Q$ from $\Vin$ to
$\Vout$.  For the sake of simplicity we assume $\Vin=\{R\}$ and
$\Vout=\{S\}$ consist of a single relation name.
Then $\Prog$ is a set of rules with $S$ in the head predicate and
$R$ as the only EDB relation.
\new{By Lemma~\ref{eqelim}, we can transform $\Prog$ into a 
program without equalities, so we will assume from here on that
$\Prog$ contains no equalities.}

Recall from Example~\ref{exdelta} the set of six jaegds $\Delta =
\{\delta_1,\dots,\delta_6\}$ that expresses properness.  The main
idea is that $Q$ has the object--object property if and only if
the ``$\Delta$--$\Delta$ implication problem'' holds for $Q$.
We then leverage the observation made in
Example~\ref{exdeltacomplete} that chasing from $\Delta$ is
unambiguous.

More precisely, for a relation name $P$ and each $i=1,\dots,6$,
let $\delta_i^P$ be the version of $\delta_i$ where we substitute
$P$ for the name $D$.  Let $\Delta^R =
\{\delta_1^R,\dots,\delta_6^R\}$.  Then for each each
$i=1,\dots,6$ and every instance $I \models \Delta^R$, we want
to check that $Q(I) \models \delta_i^S$.

Let us begin with $\delta_1^S$.  
We consider every pair of rules $(\ru_1,\ru_2)$ from
$\Prog$, where $\ru_1$ and $\ru_2$ can also be the same rule.
Let the head of $\ru_j$ be $S(e_j:t_j)$, for $j=1,2$.
We apply a variable renaming $\rho$ so that $\ru_1$ and
$\rho(\ru_2)$ have no variables in common.
Now construct a jaegd with equalities from $\delta_1$, $\ru_1$
and $\rho(\ru_2)$ as
follows.  Using fresh variables $\$x$, $@i$ and $@j$,
the body consists of the bodies of $\ru_1$ and
$\rho(\ru_2)$, together with the equalities $\$x=e_1$,
$\$x=\rho(e_2)$, $t_1=@i$, and $\rho(t_2)=@j$.  The head is
$(@i=@j)$. 

\new{Since the rules in $\Prog$ contain no equalities, and 
every head contains each variable at most once, it follows
that the constructed rule is equationally acyclic.
I follows by Lemma~\ref{eqelim},} that this jaegd 
with equalities
is equivalent to a finite set of jaegds, which we denote by
$\Delta_1^{\ru_1,\ru_2}$.  It is now clear that $Q(I) \models
\delta_1^S$ for every $I \models \Delta^R$, if and only if every
jaegd in $\Delta_1^{\ru_1,\ru_2}$ is logically implied by
$\Delta^R$.  This is a unambiguous case
of the implication problem, so it can be solved by the chase.

Checking implication for $\delta_2$--$\delta_5$ is similar.  For example,
from $\delta_3$, $\ru_1$ and $\rho(\ru_2)$ and fresh variables
$\$x$, $\$y$ and $@i$, we construct a denial constraint with
equalities having as body the bodies of $\ru_1$ and $\rho(\ru_2)$
together with the equalities $e_1=\$x$, $t_1=@i$, and
$\rho(e_2)=\$x.\$y$.
\end{proof}

\paragraph{Computational complexity} Like the implication problem
for egds in the relational model, the computational complexity of
the unambiguous cases of the implication problem for jaegds is
NP-complete.  Note, however, that in the above proof we only
need to chase jaegds from $\Delta_i^{\ru_1,\ru_2}$ with the fixed
set of jaegds $\Delta^R$.  Hence each application of the chase
would be polynomial, were it not for the following caveat.  The
caveat is that $\Delta_i^{\ru_1,\ru_2}$ is obtained after
elimination of equalities, which can result in exponentially many
rules, and these rules may be exponential in size due to the
repeated doubling.  Even when the given program has no
equalities, there are still equalities to be eliminated in the
jaegd constructed from $\ru_1$ and $\ru_2$.  We thus can only
conclude an exponential-time upper bound on the complexity of the
object--object problem for positive nonrecursive J-Logic
programs.  We leave the exact complexity open.

\section{The containment problem over flat instances}
\label{secontain}

Let $\Prog_1$ and $\Prog_2$ be J-Logic programs both expressing a
query from $\Vin$ to $\Vout$; let $Q_j$ be the query expressed by
$\Prog_j$.

Let $\F$ be a family of instances.  The \emph{containment problem
over $\F$} asks, given $\Prog_1$, $\Prog_2$, $\Vin$ and $\Vout$
as above, whether $Q_1(I) \subseteq Q_2(I)$ for all instances $I$
over $\Vin$ belonging to $\F$.  Recall that an instance is
\emph{flat} if no packed keys occur in it.  In this section we show:

\begin{theorem}
Let $\F$ be the set of flat instances, and let $\PF$ be the set
of proper flat instances.  For positive nonrecursive programs,
containment over $\F$ is decidable, and so is containment over
$\PF$.
\end{theorem}

Note that we restrict attention to flat instances.  Indeed, our
current solution does not work with packed keys in the inputs
(see Remark~\ref{rempackedvariants}).  It is an interesting topic
for further research to see whether our our solution can be
extended in the presence of packing.

To solve the containment problem over $\F$ one can take
inspiration from the inclusion problem for \emph{pattern
languages} over an infinite alphabet \cite{fr_patterninclusion}.
The main additional aspect here is the distinction between atomic
variables and path variables.

In the field of pattern languages,
a pattern is a finite sequence of constants and path variables,
so, in our terminology, a path expression without atomic
variables and packed key expressions.  The \emph{language}
of a pattern $e$ is the set $L(e)$ of all flat paths $p$ for
which there exists a valuation $h$ such that $h(e)=p$.  Here, a
flat path is a path in which no packed keys occur, i.e., a
nonempty sequence of atomic keys.  Note that this essentially
interprets patterns over an infinite alphabet, since our universe
$\dom$ of atomic keys is infinite.  In this case it is known
\cite{fr_patterninclusion} that $L(e_1) \subseteq L(e_2)$ if and
only if there exists a variable mapping $h$ such that
$h(e_2)=e_1$.  When
atomic variables come into play, however, this ``homomorphism
property'' is no longer necessary for containment.

\begin{example}
Let us allow atomic variables in patterns.  Then
consider the following four patterns:
\begin{align*}
e_1 &= \$x.\$y &
e_3 &= @x.\$y.@z \\
e_2 &= @x.\$y &
e_4 &= \$u.@v.\$w
\end{align*}
Then $e_1$ and $e_2$ describe the same language, namely all 
flat paths of length at least two.  There is a variable mapping
from $e_1$ to $e_2$ but not from $e_2$ to $e_1$, since a variable
mapping cannot map an atomic variable (in this case $@x$) to a
path variable (in this case $\$x$).  Also $e_3$ and $e_4$
describe the same language, namely all flat paths of length at
least three.  Here there is neither a variable mapping from $e_3$
to $e_4$ nor one from $e_4$ to $e_3$.

The simplistic idea to just allow weak variable mappings does not
work.  For example, there is a weak variable mapping from
$@x$ to $\$x$ but $L(\$x)$ is not contained in $L(@x)$.\qed
\end{example}

We next develop our general solution to the containment problem
over flat instances.  For simplicity, we always consider positive
nonrecursive programs expressing a query from $\Vin$ to $\Vout$
where $\Vout = \{S\}$ is a single-relation vocabulary.  Such
programs can be written as finite sets of rules with $S$ in the
head predicate and relation names from $\Vin$ in the bodies.

It is sufficient to solve the containment problem given
programs $\Prog_1$ and $\Prog_2$ where $\Prog_1$
consists of a single rule $\ru_1$ (since otherwies we can
check containment for all rules of $\Prog_1$ separately).
Moreover, we can make the following proviso:
\newtheorem*{proviso}{Proviso}
\begin{proviso}
The body of $\ru_1$ and the bodies of rules in
$\Prog_2$ do not have equalities.
Moreover, these bodies are flat, i.e., do not use packing.
\end{proviso}
The first part of the proviso is justified by Lemma~\ref{eqelim}.
The second part is justified because
we work over flat instances: non-flat bodies can
never match anyway.  The heads may still use packing.

We begin by noting that there \emph{is} a simple homomorphism
theorem when $\ru_1$ does not have path variables.
\begin{proposition} \label{prophomonopath}
Assume $\ru_1$ does not have path variables.
Then $\ru_1$ is contained in $\Prog_2$ over $\F$ if and only if
there exists a rule $\ru_2 \in
\Prog_2$ such that there is a homomorphism from $B_2$ to
$B_1$, mapping $H_2$ to $H_1$.  Here, $B_i$ and $H_i$ denote the
body and the head of $\ru_i$.
\end{proposition}
\begin{proof}
The if-direction is straightforward.  For the only-if direction,
we view $B_1$ as a (flat) instance $I$ by viewing each variable as
an atomic key (called a frozen variable).
We similarly view $H_1$ as a fact.
Then clearly $H_1 \in \ru_1(I)$, so also $H_1 \in \Prog_2(I)$.  Hence
there exists $\ru_2 \in \Prog_2$ and a valuation $\nu$ such that
$\nu(B_2) \subseteq B_1$ and $\nu(H_2)=H_1$.  Since $\ru_1$ does
not have path variables, $\nu$ can map atomic variables only to
constants or to (frozen) atomic variables.  Hence, we can view $\nu$
as a homomorphism from $\ru_2$ to $\ru_1$.
\end{proof}

We now reduce the containment problem where $\ru_1$ has path
variables, to infinitely many calls to the containment problem
where $\ru_1$ does not have path variables.  Thereto, we
associate to every path variable $\$x$ an infinite sequence
$@x^1$, $@x^2$, \dots\ of atomic variables.  Obviously, for
distinct path variables $\$x$ and $\$y$ we assume $@x^i$ and
$@y^j$ are distinct for all $i$ and $j$.

A \emph{variant} of $\ru_1$ is a rule obtained from $\ru_1$ as
follows.  For every path variable $\$x$ in $\ru_1$, choose a
natural number $n_{\$x}$.  We call $n_{\$x}$ the \emph{chosen
length} for $\$x$.  Now replace each occurrence of $\$x$
in $\ru_1$ by the sequence $@x^1\dots @x^{n_{\$x}}$.
Thus, as soon as $\ru_1$ has at least one path variable,
there are infinitely many variants of $\ru_1$.

The following is now clear:
\begin{proposition} \label{propvariants}
$\ru_1$ is equivalent, over $\F$, to the infinite union of its variants.
In particular,
$\ru_1$ is contained in $\Prog_2$ over $\F$ if and only if every
variant of $\ru_1$ is contained in $\Prog_2$ over $\F$.
\end{proposition}

\begin{remark} \label{rempackedvariants}
The above proposition only works over flat instances.  Consider,
for example, the rules
\begin{align*}
\ru_2 &= S(c:\emptyset) \gets R(@u.\$z:\emptyset) \\
\ru_1 &= S(c:\emptyset) \gets R(\$x.\$y:\emptyset)
\end{align*}
Rule $\ru_2$ tests if $R$ contains a path of length
at least two, starting with an atomic key, and with the empty
value at the leaf.  If so, the fact $S(c:\emptyset)$ is returned
($c$ is some constant). An example of a
variant of $\ru_1$, with $2$ as chosen length for $\$x$ and $3$
for $\$y$, is
$$
S(c:\emptyset) \gets R(@x^1.@x^2.@y^1.@y^2.@y^3:\emptyset).
$$
We see that this variant, and indeed every variant, of $\ru_1$ is
contained in $\ru_2$.  Nevertheless $\ru_1$ is not contained in
$\ru_2$ over all instances, as witnessed by the instance
$I=\{R(\langle a \rangle. b:\emptyset)\}$.\qed
\end{remark}

The above proposition gives us infinitely many variant
containments to check.  Our final step reduces this to a finite
number.
\begin{proposition} \label{propclb}
Let $m$ be the number of atomic variables used in $\Prog_2$.
Assume all variants of $\ru_1$ with chosen lengths up to $m+1$
are contained in $\Prog_2$ over $\F$.  Then every variant of
$\ru_1$ is contained in $\Prog_2$ over $\F$.
\end{proposition}
\begin{proof}
By Proposition~\ref{prophomonopath}, it is sufficient to show the
following claim.
\emph{Let $\ru$ be a variant of $\ru_1$ with a chosen length $k\geq m+1$
for some path variable $\$x$.  Assume there is a homomorphism $h$
from a rule $\ru_2 \in \Prog_2$ to $\ru$.
Let $\ru'$ be the same variant as $\ru$,
except that the chosen length for $\$x$ is increased to $k'>k$.
Then there is still a homomorphism from $\ru_2$ to $\ru'$.}

We argue the claim as follows.  Since $k > m$, one
of the variant variables for $\$x$, say
$@x^j$, is not in the image of $h$ applied to any atomic variable
from $\ru_2$.  Hence it only occurs in the images of some path
variables.  Each such path variable is mapped by $h$ to a flat
path expression in which $@x^j$ occurs.  Now for each such
variable $\$z$, modify $h(\$z)$ by inserting the sequence
$@x^{k+1}\dots @x^{k'}$ behind each occurrence of $@x^j$.  The
resulting variable mapping $h'$ gives us the desired
homomorphism from $\ru_2$ to $\ru'$.  (The only detail is that
the variant sequence for $\$x$ is permuted a bit, instead of
$@x^1\dots @x^{k'}$ it is now $@x^1\dots @x^j
@x^{k+1}\dots @x^{k'} @x^{j+1}\dots @x^k$.)
\end{proof}

We conclude that containment of $\ru_1$ in $\Prog_2$ over $\F$ is
decidable with $\Pi^{\rm P}_2$ complexity.  Indeed, instead of
trying all variants of $\ru_1$, as given by
Proposition~\ref{propvariants}, it suffices to try all variants
of $\ru_1$ with chosen length bounds as given by
Proposition~\ref{propclb}.  For each variant we test the
existence of a homomorphism as given by
Proposition~\ref{prophomonopath}.  We leave open whether the
problem is actually $\Pi^{\rm P}_2$-hard.

\paragraph{Containment over proper flat instances}

For simplicity, let us assume that $\Vin$ consists of a single
relation name $D$.  Recall from Example~\ref{exdelta} the set
$\Delta$ of jaegds that expresses properness. We can
chase a rule with $\Delta$ in much the same way as we chase a
jaegd as defined in Section~\ref{sect:object-object}.  We establish:
\begin{proposition}
$\ru_1$ is contained in $\Prog_2$ over $\PF$ if and only if either
chasing $\ru_1$ with $\Delta$ fails, or it succeeds and results in a rule
$\ru$ such that $\ru$ is contained in $\Prog_2$ over $\F$.
\end{proposition}
\begin{proof}
For the if-direction, first assume the chase fails.  Then
$\ru_1(I)$ is empty on all proper instances so containment holds
trivially.  Next assume the chase succeeds and results in the
rule $\ru$.  Let $I$ be a proper flat instance.  By
Lemma~\ref{lemsoundchase}, appropriately adapted to rules, we
have $\ru_1(I) = \ru(I)$.  By the given, $\ru(I) \subseteq
\Prog_2(I)$ and we are done.

For the only-if direction,
suppose the chase succeeds and
results in the rule $\ru = H \gets B$.  By
Proposition~\ref{propvariants}, we have to show that
every variant $\ru'=H'\gets B'$
of $\ru$ is contained in $\Prog_2$.
We can view $B$ as an
instance $I$ by viewing each variable as an atomic key ($I$ is
called a frozen body).  Because chasing from $\Delta$ is unambiguous,
we obtain as in the proof of Theorem~\ref{propchasecomplete}
that $I \models \Delta$, i.e., $I$ is proper.

But then $I'$, the frozen variant body $B'$, is also proper.
Indeed, by replacing each frozen path variable by a sequence of
frozen atomic variables, the functional dependency from paths to
atomic values remains satisfied.  Moreover, $I'$ is also still
prefix-free.  In proof, suppose $D(p':v) \in I'$.  The last
element $s$ of $p'$ is either a constant or a frozen atomic
variable from $\ru$, or a frozen atomic variable $@x^n$ coming
from a path variable $\$x$ in $\ru$.  In the latter case, $n$
must be the chosen length for $\$x$.  Now suppose there would
exists $D(p'.q':u) \in I'$.  Since the first symbol of $q'$
follows the last symbol of $p'$, it is either
again a constant or frozen atomic variable from $\ru$, 
or a frozen atomic variable $y^1$ coming from a path variable
$\$u$ in $\ru$.  We conclude that the presence of $D(p':v)$ and
$D(p'.q':u)$ in $I'$ would imply the presence of some $D(p:v)$
and $D(p.q:u)$ in $I$, which is impossible because $I$ is
prefix-free.

Clearly, $H' \in \ru'(I')$, so also $H' \in \ru(I')$ since $\ru'$
is a variant of $\ru$.  Furthermore, since $\ru$ was obtained from $\ru_1$ by
applying chase steps, which are applications of homomorphisms,
also $H' \in \ru_1(I')$.  By the given, then $H' \in \Prog_2(I')$.
This means there exists $\ru_2 \in \Prog_2$ and a
matching $\nu:B_2 \to I'$ such that $\nu(B_2) \subseteq I'$ and
$\nu(H_2)=H'$.  Since $\ru'$ does not have path variables, $\nu$
can be viewed as a homomorphism from $\ru_2$ to $\ru'$.  Hence,
$\ru'$ is contained in $\Prog_2$ as desired.
\end{proof}

\section{Conclusion} \label{seconc}

Thanks to the deterministic nature of JSON objects, it is very
convenient to view objects as sets of key sequences paired with
atomic values.  We recommend the use of path variables, ranging
over key sequences, in languages for JSON querying for accessing
deeply nested data.  While the data complexity is
polynomial-time, it would be interesting to investigate practical
query processing issues involving path variables.

Furthermore, packing is a versatile tool not
only for expressive power and the generation of new keys, but
also for marking parts of sequences, duplicate elimination, and
other tricks.  We recommend that practical JSON query processors
support packed keys.

Our technical results have shown that the proposed approach is
workable.  Much further work can be done: Is there a nonrecursive
flat--flat theorem? What is the exact complexity of the
object--object problem for nonrecursive programs? Is the
containment problem for nonrecursive programs decidable in the
presence of packing?  How does the complexity of the containment
problem change when equalities are allowed in rules?

During our research we also encountered the following intriguing
puzzle.  Consider the extreme case where there exists only one
atomic key, and there is no packing.  Then J-Logic amounts to
monadic Datalog with stratified negation over sets of sequences
of $a$'s, with path variables and atomic variables.  This
corresponds to monadic Datalog with stratified negation over sets
of natural numbers, with natural number constants and variables,
and addition as the only operation.  Which functions on sets of
natural numbers are expressible in this language?

\paragraph{Acknowledgment}

We thank Dominik Freydenberger and Georg Gottlob for helpful
communications.

\bibliographystyle{plain}
\bibliography{database}

\end{document}